\documentclass[a4paper,twocolumn,english,showpacs]{revtex4}
\usepackage[T1]{fontenc}
\usepackage[latin9]{inputenc}
\usepackage{float}
\usepackage{amsmath}
\usepackage{graphicx}
\usepackage{esint}



\usepackage{babel}

\begin{document}

\title{Energy harvesting by utilization of nanohelices }

\author{Sung Nae Cho  }

\email{sungnae.cho@samsung.com}

\affiliation{MEMS \& Packaging Group, Samsung Advanced Institute of Technology,
Mt. 14-1 Nongseo-dong Giheung-gu, Yongin-si Gyeonggi-do, 446-712,
South Korea }

\date{Revised 18 March  2009}

\begin{abstract}
An energy harvesting device based on nanohelices is presented. The
energy harvesting scheme based on nanohelices involves the same rectification
circuitry found in many household electronic goods, which converts
alternating current (\textbf{AC}) from a wall outlet into a direct
current (\textbf{DC}) supply. The presented device, however, involves
the rectification of ambient electromagnetic waves rather than the
\textbf{AC} source from a household wall outlet. 
\end{abstract}


\maketitle

\section{Introduction}

Antennas are common in many wireless devices, such as cordless phones,
radios, and television sets. For radio and telecommunication applications,
antennas are designed to receive electromagnetic waves in the gigahertz
frequency ranges. The electromagnetic signals received are converted
into electrical currents, which in turn generate sound, images, and
so on depending on the type of device. In principle, antennas are
the most fundamental energy harvesting devices. 

The idea of collecting solar energy by antenna dates back as early
as 1970's\citet{Bailey}. Since then, researches in solar energy collection
by antennas have slowly progressed\citet{Corkish-2002,Corkish-2003,Karmakar,Hudak}.
However, due to the limitations on physical size of antenna, it was
only recently a significant achievement in energy harvesting by antennas
has been realized for the infrared (\textbf{IR}) spectrum of electromagnetic
waves\citet{Kotter}. For telecommunication applications, dimension
of antenna is on the order of centimeters. For the \textbf{IR} spectrum
of electromagnetic energies, the antenna size scales on the order
of sub-microns and this makes harvesting energy from light by antenna
even more challenging. 

The efficiency of an antenna strongly depends on its size\citet{Balanis}.
With the advent of nanotechnology, the abundance of sub-micron sized
structures which can be used as antenna exists today. Nano structures,
such as nanorods, nanotubes, and nanodots, are beginning to shed some
light on harvesting energy from electromagnetic radiation in \textbf{IR}
to ultraviolet (\textbf{UV}) spectrum of range\citet{Tsakalakos,Saychev,Chen}.
The size (or dimension) is not the only physical property that affects
the efficiency of an antenna. For more sophisticated antennas, its
geometrical configuration, e.g., shape, significantly affects the
antenna efficiency\citet{Balanis}. Antennas based on simple nanorods,
nanotubes, or nanodots leave little room for manipulating their geometrical
configurations for optimizations. This put helical antennas out of
the picture for developing solar cells based on antenna theory. This
is all about to change with recent developments in nanohelices\citet{D. Zhang-nanospring-1,Nakamatsu-nanospring-2,Dice-nanospring-VG-1,Daraio-nanospring-VG-2,Zhang-nanospring-VG-3,Singh-nanospring-VG-4}.

\begin{figure}[H]
\begin{centering}
\includegraphics[scale=0.6]{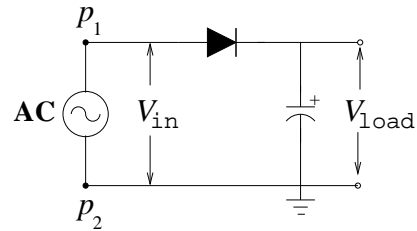}
\par\end{centering}

\caption{\label{HWR-with-CAP}Half-wave rectifier with capacitor.}

\end{figure}

The helical antennas are widely deployed technology, which is well
documented and studied in literature\citet{Phillips-Helical-antenna-1,Nakano-helical-antenna-2}.
Perhaps, the most widely deployed helical antenna, but which is also
least likely to be thought of as one, i.e., as an helical antenna,
is the transformer found in many electronic appliances. Because majority
of battery operated electronic devices run on direct current (\textbf{DC})
power, the electromagnetic energy harvested by an antenna, which is
an alternating current (\textbf{AC}) power, must be rectified. The
circuitry that rectifies an \textbf{AC} into a \textbf{DC} power is
referred to as a rectifier and one of the breed of rectifiers, a half-wave
rectifier, is illustrated in Fig. \ref{HWR-with-CAP}. 

\begin{figure}[H]
\begin{centering}
\includegraphics[scale=0.6]{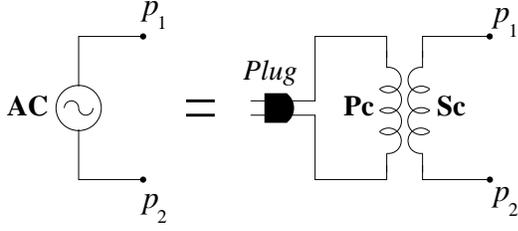}
\par\end{centering}

\caption{\label{transformer} Representation of a transformer as the \textbf{AC}
source. }

\end{figure}

The \textbf{AC} source in the rectifier circuit is physically represented
by a transformer, Fig. \ref{transformer}. The primary and secondary
coils are labeled as \textbf{Pc} and \textbf{Sc}, respectively. Basically,
transformer is an union of two helical conductors in close proximity,
where one of them is termed primary coil and the other is termed secondary
coil. Making analogy with the radio station and a radio, the primary
coil plays the role of radio station and the secondary coil plays
the role of a radio. In a transformer, primary coil transmits electromagnetic
waves and the secondary coil receives those waves and converts them
into electrical currents. 

The term rectification becomes meaningless unless the output voltage,
$V_{\textup{load}},$ and the input voltage, $V_{\textup{in}},$ satisfy
the rectification condition, \begin{align}
V_{\textup{in}} & \geq V_{\textup{load}}+V_{\textup{be}},\label{eq:rectification_condition}\end{align}
 where $V_{\textup{be}}$ is the voltage drop across the diode and
$V_{\textup{in}}$ is the voltage amplitude of the \textbf{AC} source%
\footnote{For the full-wave bridge rectification, the $V_{\textup{be}}$ is
replaced by $2V_{\textup{be}}.$%
}. The smaller magnitude for $V_{\textup{be}}$ translates into higher
system efficiency. For normal diodes, $V_{\textup{be}}$ ranges between
$0.6\sim1.8$ volts. Schottky diode, which is a special type of diode
with very low forward-voltage drop, has the $V_{\textup{be}}$ between
approximately $0.1\sim0.5$ volts. 

\begin{figure}[H]
\begin{centering}
\includegraphics[scale=0.6]{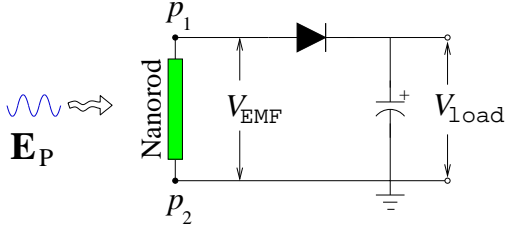}
\par\end{centering}

\caption{\label{nanorod-antenna} Simple nanorod rectenna, i.e., {}``rectifying
antenna.'' }

\end{figure}

The household wall outlet supplies anywhere from $115\,\textup{V}$
to $220\,\textup{V}$ for $V_{\textup{in}},$ which is much larger
than the diode forward voltage drop.  Therefore, $V_{\textup{be}}$
of $0.6\sim1.8$ volts is not of much concern when rectifying power
line voltage. But, is this the same case for rectification involving
ambient electromagnetic waves? To answer this, I shall consider a
simple rectenna illustrated in Fig. \ref{nanorod-antenna}.

\begin{figure}[H]
\begin{centering}
\includegraphics[scale=0.6]{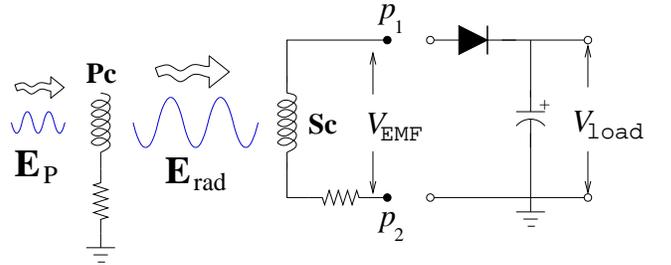}
\par\end{centering}

\caption{\label{helix-helix-amplification} Amplification of $V_{\textup{EMF}}$
by secondary radiation process. }

\end{figure}

An ambient electromagnetic plane wave with $E_{\textup{P}}\equiv\left\Vert \mathbf{E}_{\textup{P}}\right\Vert $
for the magnitude of its electric field part produces intensity $I$
given by \begin{eqnarray*}
I=\frac{1}{2}c\varepsilon_{0}E_{\textup{P}}^{2}, &  & \left\{ \begin{array}{c}
c\approx3\times10^{8}\,\textup{m}\,\textup{s}^{-1},\qquad\qquad\quad\;\:\\
\varepsilon_{0}\approx8.9\times10^{-12}\,\textnormal{\textup{s}}^{4}\,\textnormal{\textup{A}}^{2}\,\textnormal{\textup{m}}^{-3}\,\textnormal{\textup{kg}}^{-1},\end{array}\right.\end{eqnarray*}
 where $c$ is the speed of light in vacuum, $\varepsilon_{0}$ is
the permittivity of free space, and $\mathbf{E}_{\textup{\textnormal{P}}}$
is the electric field\citet{Griffiths}. In general, the radiation
from sun or nearby heat source does not form plane waves. However,
if the longest dimension of the nanorod is comparable to the wavelength
of incidence wave, the incidence wave can be approximated as a plane
wave\citet{Max Born}. The intensity of one watt per squared meters
corresponds to the electric field magnitude of \begin{eqnarray*}
I=1\,\textup{W}\,\textup{m}^{-2}, &  & E_{\textup{P}}\approx2.7\times10^{-8}\,\textup{V}\,\textup{nm}^{-2}.\end{eqnarray*}
 For the nanorod of length $l,$ the electromotive force (\textbf{EMF})
generated inside of it would be given by \begin{align*}
V_{\textup{EMF}} & =\int_{0}^{l}\mathbf{E}_{\textup{\textnormal{P}}}\cdot\mathbf{dl}=E_{\textup{P}}l,\end{align*}
 where, for simplicity, $\mathbf{E}_{\textup{P}}$ has been assumed
to be parallel to the length of nanorod. For the nanorod of $1\,\textup{um}$
in length, intensity of $1\,\textup{W}\,\textup{m}^{-2}$ generates
\[
V_{\textup{EMF}}\approx2.7\times10^{-5}\,\textup{V}\]
 inside the nanorod. The $V_{\textup{EMF}}$ generated inside the
nanorod rectenna can be identified with the $V_{\textup{in}}$ of
Fig. \ref{HWR-with-CAP} and the rectification condition, Eq. (\ref{eq:rectification_condition}),
gives \[
2.7\times10^{-5}\,\textup{V}\geq V_{\textup{load}}+V_{\textup{be}}.\]
 But, this cannot be satisfied for any $V_{\textup{load}}\geq0$ even
with a Schottky diode, which is known to have very low forward-voltage
drop. Can $V_{\textup{EMF}}$ be amplified so that the rectification
condition, Eq. (\ref{eq:rectification_condition}), is satisfied for
sufficiently large $V_{\textup{load}}?$ The answer is yes and this
involves the secondary radiation process, which is illustrated in
Fig. \ref{helix-helix-amplification}. 

The irradiance from ambient source, which is indicated by $\mathbf{E}_{\textup{\textnormal{P}}}$
(electric field part) in the figure, induces radiation from the primary
coil, \textbf{Pc}, which is indicated by $\mathbf{E}_{\textup{\textnormal{rad}}}$
(electric field part only). Solution obtained by solving Maxwell equations
shows that $\mathbf{E}_{\textup{\textnormal{rad}}}\gg\mathbf{E}_{\textup{\textnormal{P}}}$
when the secondary coil, \textbf{Sc}, is placed very close to the
primary coil. Since $\mathbf{E}_{\textup{\textnormal{rad}}}$ acts
as the incidence wave for \textbf{Sc}, the $V_{\textup{EMF}}$ is
amplified in \textbf{Sc} by factor of $\mathbf{E}_{\textup{\textnormal{rad}}}/\mathbf{E}_{\textup{\textnormal{P}}}.$

\begin{figure}[H]
\begin{centering}
\includegraphics[scale=0.5]{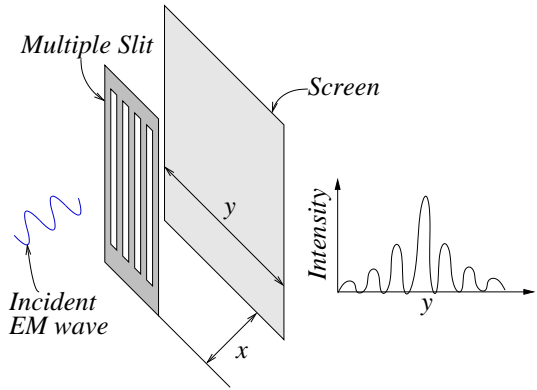}
\par\end{centering}

\caption{\label{multi-slit}Multiple-slit diffraction.}

\end{figure}

The amplification of $V_{\textup{EMF}}$ by secondary radiation process
can be qualitatively understood by recalling the multiple-slit experiment
with coherent light source, Fig. \ref{multi-slit}. When plane waves
pass through a multiple-slit plate, at distance $x$ away from the
slits, wavelets couple either constructively or destructively depending
on the location of $y$ and this results in bright and dark intensity
patterns on screen. The cross-section of nanohelix, which forms the
primary and the secondary coils in Fig. \ref{helix-helix-amplification},
resembles the multiple-slit plate (except here, the slit pattern is
in ordered zigzag form).

That being said, the plane wave condition and the coherence of incidence
wave is crucial to the amplification of $V_{\textup{EMF}}$ by secondary
radiation process. The outdoor sun-light or the irradiance from light-bulb
are not plane waves if plane waves are thought of as wave front with
definite degree of coherence,  which can be measured by the visibility
of interference\citet{Max Born}. To put it simply, the degree of
coherence is a measure of how perfectly the waves can cancel due to
destructive interference (or the opposite, measure of how perfectly
the waves can add up due to constructive interference). The coherence
was originally introduced in connection with Young's double-slit experiment
in optics, where the interference becomes visible when light is allowed
to pass through small aperture such as pin-hole and the effect becomes
more pronounced with smaller pin-holes regardless of the light source.
Young's experiment justifies the use of plane wave input for nanohelices
considered here as its height and winding pitch scales on the order
of wavelength%
\footnote{The winding pitch for the typical nanohelices are several orders or
more smaller than the wavelength of the incidence wave.%
}.

\section{Nanotransformer energy harvesting device}

\subsection{Device structure}

\begin{figure}[H]
\begin{centering}
\includegraphics[scale=0.5]{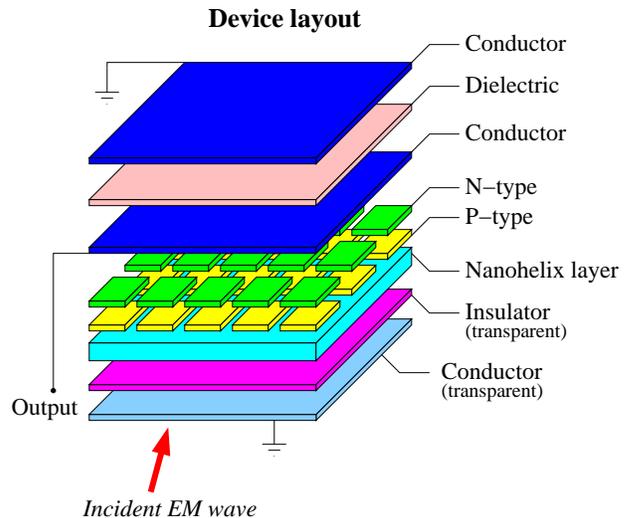}
\par\end{centering}

\caption{\label{NEH_device} Nanotransformer energy harvesting device layout. }

\end{figure}

\begin{figure}[H]
\begin{centering}
\includegraphics[scale=0.5]{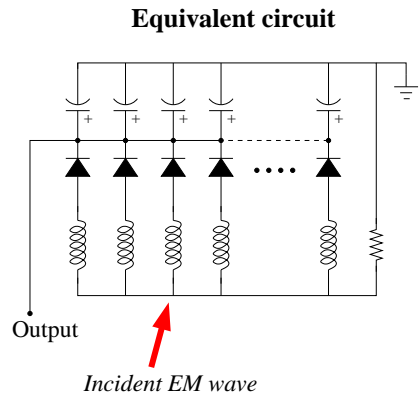}
\par\end{centering}

\caption{\label{NEH_equivalent_circuitry} Equivalent circuitry.}

\end{figure}

\begin{figure}[H]
\begin{centering}
\includegraphics[scale=0.5]{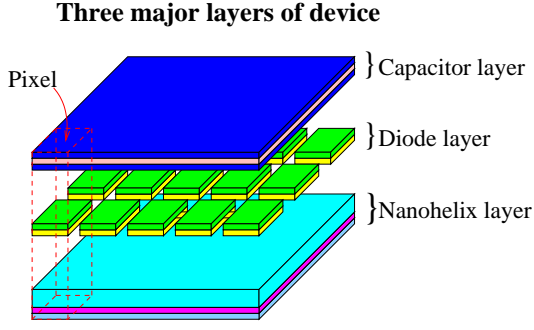}
\par\end{centering}

\caption{\label{NEH_device_stage} Three major layers of nanotransformer energy
harvesting device. }

\end{figure}

The physical layout of energy harvesting device based on nanohelices
(or nanotransformer\citet{Scho-patent-1,Scho-patent-2}) is illustrated
in Fig. \ref{NEH_device} and its equivalent circuit diagram is provided
in Fig. \ref{NEH_equivalent_circuitry}. Borrowing the terminology
from display technology, I shall refer to each element in diode layer
(indicated by N-type and P-type square pairs in Fig. \ref{NEH_device})
and nanohelices making contact with the diode element as a pixel.
For clarity, the pixel is indicated by dotted rectangular three dimensional
cube in Fig. \ref{NEH_device_stage}. 

The proposed energy harvesting device based on nanohelices involves
three major device layers (capacitor, diode, and nanohelix layers),
which can be processed independently and sandwiched together for the
final product, Fig. \ref{NEH_device_stage}. Such process allows manufacturing
of the proposed energy harvesting device at large scales. 

\begin{figure}[H]
\begin{centering}
\includegraphics[scale=0.5]{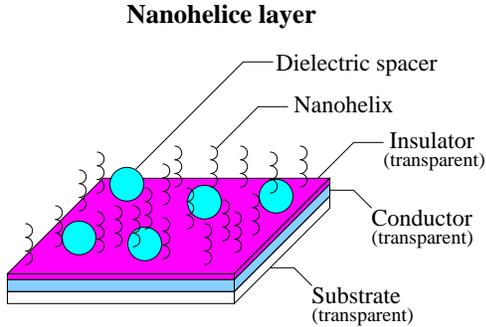}
\par\end{centering}

\caption{\label{Nanohelice_layer} Schematic of grown nanohelices.}

\end{figure}

\begin{figure}[H]
\begin{centering}
\includegraphics[scale=0.5]{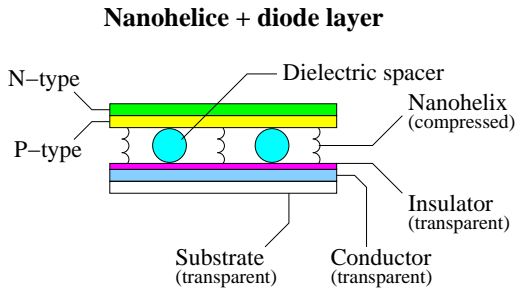}
\par\end{centering}

\caption{\label{helice_plus_diode_layer} The cross-sectional view of nanohelix
layer with diode layer placed over it.}

\end{figure}

The diode layer can be first prepared as a one large sheet of a diode,
which then can be patterned to form pixels of diodes. The nanohelix
layer in current device scheme can be prepared by growing helices
in normal direction of the substrate as illustrated in Fig. \ref{Nanohelice_layer}.
Equivalently, nanohelices can also be spin coated directly onto the
surface of a substrate. In this case, nanohelices would most likely
be positioned with its length parallel to the surface of a substrate.
To prevent diode layer from collapsing onto the substrate of nanohelix
layer, transparent dielectric spacer such as $\textup{Si}\textup{O}_{2}$
nanoparticles may be distributed on the substrate surface as illustrated
in Figs. \ref{Nanohelice_layer} and \ref{helice_plus_diode_layer}.

\subsection{Operation principle}

For a successful rectification, the $V_{\textup{EMF}}$ generated
across each pixel must be amplified large enough to satisfy the rectification
condition, Eq. (\ref{eq:rectification_condition}). As already discussed
in previous sections, the amplification of $V_{\textup{EMF}}$ is
achieved thru the process of secondary amplification, recall Fig.
\ref{helix-helix-amplification}. The role of primary coil, i.e.,
\textbf{Pc} in Fig. \ref{helix-helix-amplification}, in current device
scheme, Figs. \ref{NEH_device} and \ref{NEH_equivalent_circuitry},
is played out by nanohelices belonging to the neighboring pixels.
Due to constructive and destructive interferences of wavelets originating
from different nanohelices, there would be pixels receiving amplified
radiation fields and there would be those pixels receiving virtually
no radiation fields. Only those pixels positioned in locations where
constructive interference occurs would generate large enough $V_{\textup{EMF}}$
to meet the rectification condition and, eventually, participate in
energy harvesting. 

The number of pixels, where a pixel is indicated by dotted rectangular
three dimensional cube in Fig. \ref{NEH_device_stage}, plays the
key role in the energy harvesting based on nanohelices. As an illustration,
assume that each pixel has a dimension of $10\,\textup{um}\times10\,\textup{um}$
for its surface area. In an ideal close packing, about  $100$ million
such pixels would be able to fit in an area of $10\,\textup{cm}\times10\,\textup{cm}.$
Weber et al. have experimentally shown that $\textup{ZnO}$ nanowire
can carry up to roughly $330\,\textup{nA}$ of electrical current
before it snaps\citet{D. H. Weber}. The current limiting resistors
in Figs. \ref{helix-helix-amplification}, \ref{NEH_device}, and
\ref{NEH_equivalent_circuitry} prevents the overloading of nanohelices
with too much current, thereby saving it from a break down. If assumed
that each pixel contains single nanohelix and that each nanohelix
carries electrical current of $130\,\textup{nA},$ this amounts to
a total of $13\,\textup{A}$ out of the device. Of course, only those
pixels positioned in locations where constructive interference occurs
to generate $V_{\textup{EMF}}$ large enough to meet the rectification
condition contribute to the total current. But, even if one assumes
that only $10\%$ of $100$ million pixels contribute in energy harvesting,
this still yields the total of $1.3\,\textup{A}$ from the device. 

Having said enough about the potential of energy harvesting based
on nanohelices, the validity of working principles behind the proposed
device depend heavily on the amplification of ambient electromagnetic
fields by secondary radiation process, thru which process generates
$V_{\textup{EMF}}$ large enough to satisfy the rectification condition.
Since the amplification of $V_{\textup{EMF}}$ depends on both magnetic
induction (~$\mathbf{B}_{\textup{\textnormal{rad}}}$~) and electric
field (~$\mathbf{E}_{\textup{\textnormal{rad}}}$~) parts of the
electromagnetic radiation from the primary helix, \[
V_{\textup{EMF}}\propto\left\{ \begin{array}{c}
\left\Vert \mathbf{E}_{\textup{\textnormal{rad}}}\right\Vert \\
\left\Vert \mathbf{B}_{\textup{\textnormal{rad}}}\right\Vert \end{array}\right.,\]
 one needs to quantitatively show that indeed $\mathbf{E}_{\textup{\textnormal{rad}}}$
and $\mathbf{B}_{\textup{\textnormal{rad}}}$ get amplified significantly,
\begin{eqnarray}
\frac{E_{\textup{rad}}}{E_{\textup{P}}}\equiv\frac{\left\Vert \mathbf{E}_{\textup{\textnormal{rad}}}\right\Vert }{\left\Vert \mathbf{E}_{\textup{\textnormal{P}}}\right\Vert }\gg1, &  & \frac{B_{\textup{rad}}}{B_{\textup{P}}}\equiv\frac{\left\Vert \mathbf{B}_{\textup{\textnormal{rad}}}\right\Vert }{\left\Vert \mathbf{B}_{\textup{\textnormal{P}}}\right\Vert }\gg1.\label{eq:amplification_proof}\end{eqnarray}
 The quantitative verification of Eq. (\ref{eq:amplification_proof})
involves the solving of Maxwell equations and this is the task which
I set out to do in the next sections.

\section{Theory}

All phenomena involving interaction with electromagnetic waves involve
Maxwell equations. To keep the topic presented here self-contained,
I shall briefly summarize the kind of manipulations and approximations
assumed in obtaining the vector potential partial differential equation
(\textbf{PDE}), which marks the starting point for the rest of analysis
throughout this presentation.

\subsection{Maxwell  equations}

Maxwell equations, in the form independent of particular system of
units, may be expressed as \begin{eqnarray*}
\nabla\cdot\mathbf{E}=4\pi g\rho, &  & \nabla\cdot\mathbf{B}=0,\\
\nabla\times\mathbf{E}=-\eta\frac{\partial\mathbf{B}}{\partial t}, &  & \nabla\times\mathbf{B}=4\pi\gamma\mathbf{J}+\frac{\gamma}{g}\frac{\partial\mathbf{E}}{\partial t},\end{eqnarray*}
 where $\rho$ is the total charge density, $\mathbf{J}$ is the total
current density, $\mathbf{B}$ is the magnetic induction, $\mathbf{E}$
is the electric field, and the positive constants $g,$ $\gamma,$
and $\eta$ depend on the particular system of units being adopted\citet{Davis-Snider}.
If one assumes  that both charge and current densities are specified
throughout space and assume that $\mathbf{E},$ $\mathbf{B},$ $\rho,$
and $\mathbf{J}$ vary in time as $\exp\left(i\omega t\right),$ Maxwell
 equations may be re-expressed in an alternate form as \begin{eqnarray*}
\nabla\cdot\mathbf{E}=4\pi g\rho, &  & \nabla\cdot\mathbf{B}=0,\\
\nabla\times\mathbf{E}=-i\omega\eta\mathbf{B}, &  & \nabla\times\mathbf{B}=4\pi\gamma\mathbf{J}+\frac{i\omega\gamma}{g}\mathbf{E},\end{eqnarray*}
 where $\omega$ is the angular frequency. For a non-static case,
where $\omega\neq0,$ the electric divergence relation, \[
\nabla\cdot\mathbf{E}=4\pi g\rho,\]
 becomes redundant%
\footnote{ The redundancy of $\nabla\cdot\mathbf{E}=4\pi g\rho$ for $\omega\neq0$
can be shown by taking the divergence of $\nabla\times\mathbf{B},$
\[
\nabla\cdot\nabla\times\mathbf{B}=4\pi\gamma\nabla\cdot\mathbf{J}+\frac{i\omega\gamma}{g}\nabla\cdot\mathbf{E}=0\]
 to yield  \[
\frac{i\omega\gamma}{g}\nabla\cdot\mathbf{E}=-4\pi\gamma\nabla\cdot\mathbf{J}.\]
 Finally, insertion of the continuity equation, $\nabla\cdot\mathbf{J}=-i\omega\rho,$
proves the result, \[
\frac{i\omega\gamma}{g}\nabla\cdot\mathbf{E}=i\omega\gamma4\pi\rho\Rightarrow\nabla\cdot\mathbf{E}=4\pi g\rho.\]
} and the problem of electrodynamics is reduced to solving the following
set of Maxwell equations in harmonic frequency domain, \begin{alignat}{1}
\nabla\cdot\mathbf{B} & =0,\label{eq:divB-FD}\\
\nabla\times\mathbf{E} & =-i\omega\eta\mathbf{B},\label{eq:curlE-FD}\\
\nabla\times\mathbf{B} & =4\pi\gamma\mathbf{J}+\frac{i\omega\gamma}{g}\mathbf{E}.\label{eq:curlB-FD}\end{alignat}

\subsection{Vector potential }

I proceed by seeking a vector field solution that simultaneously satisfies
Maxwell equations (\ref{eq:divB-FD}) thru (\ref{eq:curlB-FD}). Any
vector field $\mathbf{A},$ which satisfies the condition \begin{equation}
\mathbf{B}=\nabla\times\mathbf{A},\label{eq:curlA}\end{equation}
 automatically satisfies Eq. (\ref{eq:divB-FD}) and such vector $\mathbf{A}$
is given a name {}``vector potential.'' Substitution of Eq. (\ref{eq:curlA})
in Eq. (\ref{eq:curlE-FD}) gives \[
\nabla\times\left(\mathbf{E}+i\omega\eta\mathbf{A}\right)=0.\]
 The fundamental theorem of vector analysis tells us that any scalar
field $\Phi$ satisfies the condition $\nabla\times\nabla\Phi=0.$
And, this implies \[
\mathbf{E}+i\omega\eta\mathbf{A}=\nabla\Phi,\]
 where the sign of $\Phi$ is arbitrary. However, because it has already
been defined in literature that $\mathbf{E}=-\nabla\Phi$ for the
static limit, where $\omega=0,$ one chooses  $\Phi\rightarrow-\Phi$
for the scalar field and the previous relation becomes \begin{equation}
\mathbf{E}=-i\omega\eta\mathbf{A}-\nabla\Phi.\label{eq:E-in-A-Phi}\end{equation}
 Equation (\ref{eq:E-in-A-Phi}) automatically becomes the static
limit expression in the limit $\omega$ goes to zero. 

The concept of vector and scalar fields simplify electromagnetic problem
to solving of  a single Maxwell equation (\ref{eq:curlB-FD}). Insertion
of Eqs. (\ref{eq:curlA}) and (\ref{eq:E-in-A-Phi}) into Eq. (\ref{eq:curlB-FD})
gives \[
\nabla\times\nabla\times\mathbf{A}=4\pi\gamma\mathbf{J}+\frac{i\omega\gamma}{g}\left(-i\omega\eta\mathbf{A}-\nabla\Phi\right).\]
 Application of the vector identity,  \[
\nabla\times\nabla\times\mathbf{A}=-\nabla^{2}\mathbf{A}+\nabla\left(\nabla\cdot\mathbf{A}\right),\]
 transforms the previous relation as \[
\nabla^{2}\mathbf{A}-\nabla\left(\nabla\cdot\mathbf{A}\right)=-4\pi\gamma\mathbf{J}-\frac{\omega^{2}\gamma\eta}{g}\mathbf{A}+\frac{i\omega\gamma}{g}\nabla\Phi.\]
 After some rearrangements, I arrive at the expression, \begin{equation}
\nabla\left(\nabla\cdot\mathbf{A}+\frac{i\omega\gamma}{g}\Phi\right)=\nabla^{2}\mathbf{A}+\frac{\omega^{2}\gamma\eta}{g}\mathbf{A}+4\pi\gamma\mathbf{J}.\label{eq:A-J-PDE}\end{equation}
 Since any $\mathbf{A}$ and $\Phi$ satisfying Eq. (\ref{eq:curlA})
and Eq. (\ref{eq:E-in-A-Phi}), respectively, solves Eq. (\ref{eq:A-J-PDE}),
one is  free to choose any convenient $\mathbf{A}$ and $\Phi$ so
that Eq. (\ref{eq:A-J-PDE}) becomes solvable. Choosing the following
expression for $\Phi,$  \begin{equation}
\Phi=\frac{ig}{\omega\gamma}\nabla\cdot\mathbf{A},\label{eq:Cap_Phi}\end{equation}
 makes the left side of Eq. (\ref{eq:A-J-PDE}) to vanish. And, the
electric field, utilizing Eq. (\ref{eq:E-in-A-Phi}), can be expressed
as \begin{equation}
\mathbf{E}=-i\omega\eta\mathbf{A}-\frac{ig}{\omega\gamma}\nabla\left(\nabla\cdot\mathbf{A}\right).\label{eq:E-in-A}\end{equation}
 With $\Phi$ defined in Eq. (\ref{eq:Cap_Phi}), the vector potential
satisfies the following partial differential equation (\textbf{PDE}),
\begin{equation}
\nabla^{2}\mathbf{A}+g^{2}\mathbf{A}=-4\pi\gamma\mathbf{J},\quad K=\omega\sqrt{\frac{\gamma\eta}{g}}>0,\label{eq:A-PDE}\end{equation}
 where the constant $K$ has the physical implication of being the
wave number. Equation (\ref{eq:A-PDE}) is the well known Helmholtz
equation and its solution is given by \begin{equation}
\mathbf{A}\left(\mathbf{R}\right)=\gamma\iiint\frac{\mathbf{J}\exp\left(-iK\left\Vert \mathbf{R}-\mathbf{R}_{\textup{s}}\right\Vert \right)}{\left\Vert \mathbf{R}-\mathbf{R}_{\textup{s}}\right\Vert }dV_{\textup{s}},\label{eq:A-Integral}\end{equation}
 where $\mathbf{R}_{\textup{s}}$ is the position of current density,
$\mathbf{J}\equiv\mathbf{J}\left(\mathbf{R}_{\textup{s}}\right),$
and the volume integration is performed over entire region containing
the current source.

\section{Analysis}

\subsection{Nanohelix}

\begin{figure}
\begin{centering}
\includegraphics[scale=0.4]{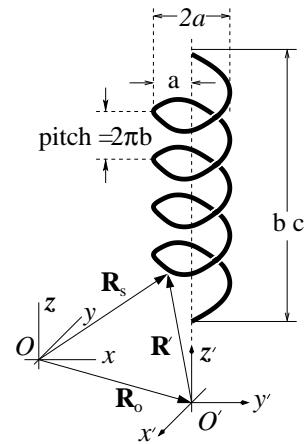}
\par\end{centering}

\caption{\label{HS-with-bc}Finite helix, where $\mathbf{e}_{1}\parallel x\parallel x^{\prime},$
$\mathbf{e}_{2}\parallel y\parallel y^{\prime},$ and $\mathbf{e}_{3}\parallel z\parallel z^{\prime}.$
The symbol $\parallel$ reads as {}``parallel to.'' }

\end{figure}

The simplest solenoid is given by a non-planar helical curve depicted
in Fig.  \ref{HS-with-bc}. If $\mathbf{e}_{1},$ $\mathbf{e}_{2},$
and $\mathbf{e}_{3}$ denote a right-handed system of mutually perpendicular
unit vectors, then every spatial points on filamentary finite helix
can be represented by \begin{equation}
\mathbf{R}_{\textup{s}}=\mathbf{R}_{\textup{o}}+\mathbf{R}^{\prime},\label{eq:Rs}\end{equation}
 where $\mathbf{R}_{\textup{o}}$ is the position vector defining
the local origin $O^{\prime}$ and $\mathbf{R}^{\prime}$ is the vector
defining the position of current source relative to $O^{\prime}.$
In Cartesian coordinates, $\mathbf{R}^{\prime}$ and $\mathbf{R}_{\textup{o}}$
are given by \begin{equation}
\left(\begin{array}{c}
\mathbf{R}^{\prime}\\
\mathbf{R}_{\textup{o}}\end{array}\right)=\sum_{i=1}^{3}\left(\begin{array}{c}
x_{i}^{\prime}\\
x_{\textup{o}i}\end{array}\right)\mathbf{e}_{i},\quad\left\{ \begin{array}{c}
x_{1}^{\prime}\equiv x^{\prime}=a\cos s,\\
x_{2}^{\prime}\equiv y^{\prime}=a\sin s,\,\\
x_{3}^{\prime}\equiv z^{\prime}=bs.\quad\;\;\:\end{array}\right.\label{eq:x-y-z-prime}\end{equation}
 For the rest of the presentation, I shall designate the trio $\left(\mathbf{e}_{1},\mathbf{e}_{2},\mathbf{e}_{3}\right)$
with $\left(x,y,z\right)$ in respective order. Similarly, for the
$O^{\prime}$ coordinates, I shall designate the trio $\left(\mathbf{e}_{1},\mathbf{e}_{2},\mathbf{e}_{3}\right)$
with $\left(x^{\prime},y^{\prime},z^{\prime}\right)$ in respective
order. The coordinates $x^{\prime}$ and $y^{\prime}$ describe a
circle of radius $a,$ and the $z^{\prime},$ which coordinate defines
the height of finite helix, increases or decreases in direct proportion
to the parameter $s.$ The vertical distance between the coils, which
is known as the pitch, equals the increase in $z^{\prime}$ as $s$
jumps by $2\pi.$ The pitch is hence given by\begin{equation}
\textnormal{pitch}=2\pi b.\label{eq:pitch}\end{equation}

Assuming $\mathbf{R}_{\textup{s}}$ is differentiable and that $\mathbf{R}_{\textup{o}}$
does not depend on parameter $s,$ the vector which is tangent to
the curve defining the finite helix is given by \begin{equation}
\mathbf{T}\equiv\frac{d\mathbf{R}_{\textup{s}}}{ds}=\sum_{i=1}^{3}T_{i}\mathbf{e}_{i},\label{eq:T}\end{equation}
where \begin{equation}
T_{1}=-a\sin s,\quad T_{2}=a\cos s,\quad T_{3}=b.\label{eq:T1-T2-T3-Def}\end{equation}
 The total length, $l,$ of filamentary finite helical curve is found
by taking the line integration of associated tangent vector with respect
to the parameter $s,$  \begin{alignat*}{1}
l & =\int\mathbf{T}\cdot d\mathbf{s}=\int\left\Vert \mathbf{T}\right\Vert ds=s\sqrt{a^{2}+b^{2}},\end{alignat*}
 where $d\mathbf{s}$ is the differential arc length of the finite
helix segment. One thus obtains the upper limit for the parameter
$s,$ \begin{equation}
c=\frac{l}{\sqrt{a^{2}+b^{2}}},\quad0<s\leq c.\label{eq:c}\end{equation}
 With the parameter $s,$ the entire height of finite helix is given
by $bc$ or $bl/\sqrt{a^{2}+b^{2}}$ as indicated in Fig.  (\ref{HS-with-bc}).

\begin{figure}

\begin{centering}
\includegraphics[scale=0.4]{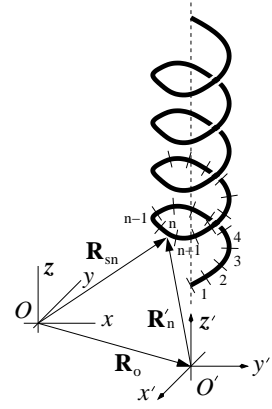}
\par\end{centering}

\caption{\label{segmented}Segmented finite helix.}

\end{figure}

The vector potential integral of Eq. (\ref{eq:A-Integral}) is integrated
over the entire volume containing the current sources. If the current
sources are confined to a filamentary finite helix whose spatial curve
is represented by Eq. (\ref{eq:Rs}), the helix may be partitioned
into segments of finite but equal sizes as illustrated in Fig.  \ref{segmented}.
The filamentary wire forming finite helix has a length of $l.$ Within
the representation parametrized in $s,$ each of $N$ segments has
length of $c/N,$ where $c$ is defined in Eq. (\ref{eq:c}). The
number of segments is arbitrary, as one can slice the wire into as
many pieces as he or she wants to. To make the argument more concise,
the filamentary wire is sliced into enough segments so that $\mathbf{J}$
for the segment is a constant and the entire segment is identified
with its center, $\mathbf{R}_{\textup{s}n},$ on the finite helix.
Then, for a detector placed at location $\mathbf{R},$ the detected
vector potential, which has been contributed from the segment $n$
on the finite helix, gets approximated by the expression  \begin{equation}
\mathbf{A}_{n}\approx\frac{\gamma\alpha_{n}c}{N}\iiint\frac{\delta\left(\mathbf{R}_{\textup{s}}-\mathbf{R}_{\textup{s}n}\right)\exp\left(-iK\left\Vert \mathbf{R}-\mathbf{R}_{\textup{s}}\right\Vert \right)}{\left\Vert \mathbf{R}-\mathbf{R}_{\textup{s}}\right\Vert }dV,\label{eq:An}\end{equation}
 where the volume integration is over all space, the quantity $c/N$
represents the length of segment, the constant $\alpha_{n}$ is related
to the local current density $\mathbf{J}_{n}$ for the $n\textnormal{th}$
segment by $\mathbf{J}_{n}=c\alpha_{n}/N,$ and $\delta\left(\mathbf{R}_{\textup{s}}-\mathbf{R}_{\textup{s}n}\right)$
is the Dirac delta function, which has the integral property, \[
\iiint f\left(\mathbf{Y}\right)\delta\left(\mathbf{Y}-\mathbf{X}\right)dV=f\left(\mathbf{X}\right).\]
 Application of the integral property of Dirac delta function on Eq.
(\ref{eq:An}) yields the result \begin{equation}
\mathbf{A}_{n}\approx\frac{\gamma\alpha_{n}c\exp\left(-iK\left\Vert \mathbf{R}-\mathbf{R}_{\textup{\textup{s}}n}\right\Vert \right)}{N\left\Vert \mathbf{R}-\mathbf{R}_{\textup{\textup{s}}n}\right\Vert }.\label{eq:An-2}\end{equation}

The detector receives contributions from all $N$ segments of the
finite helix, not just from the $n\textnormal{th}$ segment. Therefore,
summing over the contributions from all $N$ segments of the finite
helix, I have \[
\mathbf{A}\approx\sum_{n=1}^{N}\mathbf{A}_{n}=\sum_{n=1}^{N}\frac{\gamma\alpha_{n}c\exp\left(-iK\left\Vert \mathbf{R}-\mathbf{R}_{\textup{\textup{s}}n}\right\Vert \right)}{N\left\Vert \mathbf{R}-\mathbf{R}_{\textup{\textup{s}}n}\right\Vert }.\]
 Finally, in the limit the slices become finer and finer, it becomes
\[
\mathbf{A}=\lim_{N\rightarrow\infty}\sum_{n=1}^{N}\frac{\gamma\alpha_{n}c\exp\left(-iK\left\Vert \mathbf{R}-\mathbf{R}_{\textup{\textup{s}}n}\right\Vert \right)}{N\left\Vert \mathbf{R}-\mathbf{R}_{\textup{\textup{s}}n}\right\Vert }.\]

In the representation parametrized by $s,$ one notices that $c/N=\triangle s$
and, as $N$ goes to infinity, $\triangle s$ becomes infinitesimal,
i.e., $\triangle s\rightarrow ds.$ Also, as the slices get finer,
what was the center point for the $n\textnormal{th}$ slice becomes
the exact point for the slice, $\mathbf{R}_{\textup{s}n}\rightarrow\mathbf{R}_{\textup{s}}.$
Similarly, what was an average current density within the slice becomes
an exact current density for the point $\mathbf{R}_{\textup{s}},$
$\alpha_{n}\rightarrow\mathbf{J}\left(\mathbf{R}_{\textup{s}}\right).$
Hence, \[
\lim_{N\rightarrow\infty}\sum_{n=1}^{N}f\left(x\right)\frac{c}{N}\rightarrow\int f\left(x\right)ds,\]
 and the vector potential expression for the  finite helix becomes
\begin{equation}
\mathbf{A}=\gamma\int_{0}^{c}\frac{\mathbf{J}\exp\left(-iK\left\Vert \mathbf{R}-\mathbf{R}_{\textup{s}}\right\Vert \right)}{\left\Vert \mathbf{R}-\mathbf{R}_{\textup{s}}\right\Vert }ds,\label{eq:A-integral-1D}\end{equation}
 where $\mathbf{R}_{\textup{s}}$ is defined in Eq. (\ref{eq:Rs})
with its components given by Eq. (\ref{eq:x-y-z-prime}). As the current
density, $\mathbf{J},$ has not been defined, the vector potential
integral in parametrized form, Eq. (\ref{eq:A-integral-1D}), cannot
be evaluated. The current density source for the finite helix system
is cast into a quantitative form in the next section. 

\subsection{Induced current}

\begin{figure}
\begin{centering}
\includegraphics[scale=0.4]{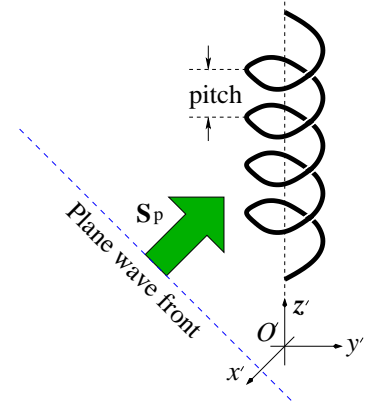}
\par\end{centering}

\caption{\label{PWIS}Plane wave incident on finite helix. The Poynting vector
is indicated by $\mathbf{S}_{\textup{p}}.$}

\end{figure}

Consider an electromagnetic problem depicted in Fig.  \ref{PWIS},
where a plane wave front is impinging on the finite helix whose configuration
is describe by Eq. (\ref{eq:Rs}). The real world solenoid, no matter
how small, always has cross-sectional area of finite size which holds
current responsible for induced electromagnetic radiation. Modeling
a real world solenoid can be difficult due to the complications arising
from a finite thickness for the cross-sectional area. However, the
mathematical modeling can be substantially simplified  by letting
the cross-sectional area of the wire to go to zero and, at the same
time, letting the current density to go to infinity in such a manner
that the flux of current along the wire remains constant. For the
current carrying helical wire modeled within such approximation, Eq.
(\ref{eq:Rs}) suffices for the description of finite solenoid. 

The vector field $\mathbf{A},$ whose solution satisfies the \textbf{PDE}
of Eq. (\ref{eq:A-PDE}), arises as a result of induced current within
the finite helix.  This induced current inside a finite helix is due
to the electric field component of impinging plane wave as illustrated
in Fig.  \ref{PWIS}. Assuming that wire forming finite helix can
be represented by an isotropic ohmic conductor, the current density
$\mathbf{J}$ is given by the Ohm's law, \begin{equation}
\mathbf{J}=\sigma\mathbf{E}_{\textnormal{\textup{p}}},\quad\left\{ \begin{array}{c}
\sigma\equiv\sigma\left(\mathbf{R}_{\textup{s}}\right),\\
\mathbf{E}_{\textnormal{\textup{p}}}\equiv\mathbf{E}_{\textnormal{\textup{p}}}\left(\mathbf{R}_{\textup{s}}\right),\end{array}\right.\label{eq:Ohms-law}\end{equation}
 where $\sigma$ is the electrical conductivity, $\mathbf{E}_{\textnormal{\textup{p}}}$
is the polarization (electric field) of impinging plane wave front,
and $\mathbf{R}_{\textup{s}}$ is the position of current source. 

\begin{figure}[H]
\begin{centering}
\includegraphics[scale=0.4]{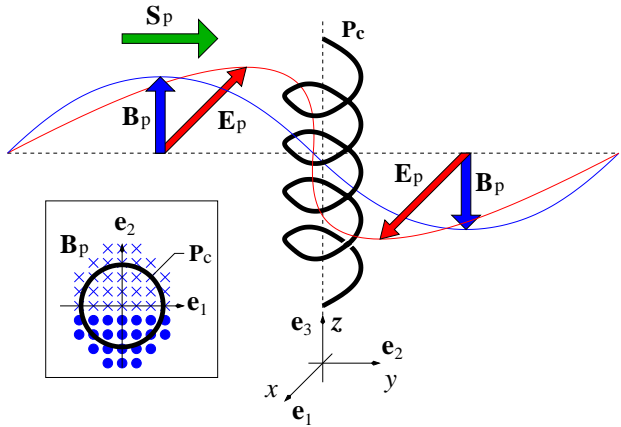}
\par\end{centering}

\caption{\label{Faraday_law_of_induction} The $\mathbf{E}_{\textnormal{\textup{p}}}$
and $\mathbf{B}_{\textnormal{\textup{p}}}$ are electric and magnetic
parts of electromagnetic wave, and $\mathbf{S}_{\textup{\textnormal{P}}}$
is the Poynting vector. The enclosed box shows the time varying flux
of magnetic induction enclosed by the nanohelix loop, where the symbol
$\times$ denotes the case of $\mathbf{B}_{\textnormal{\textup{p}}}=-B_{\textup{p}}\mathbf{e}_{3}$
and the symbol $\bullet$ denotes the case of $\mathbf{B}_{\textnormal{\textup{p}}}=B_{\textup{p}}\mathbf{e}_{3}.$}

\end{figure}

Electromagnetic waves have both electric and magnetic field parts,
as illustrated in Fig. \ref{Faraday_law_of_induction}. For a nanohelix
whose coil diameter is less than $50\,\textup{nm},$ an \textbf{IR}
range electromagnetic plane wave passing through it would be perceived
as a \textbf{DC} magnetic field switching between on and off modes
at a rate of wave frequency. This comes about because the wavelength
of electromagnetic waves in the \textbf{IR} range scales on the order
of microns, which implies the vast majority of \textbf{IR} range electromagnetic
waves are more than hundred times larger in their wavelength compared
to the $50\,\textup{nm}$ diameter of the nanohelix. And, this argument
strengthens with nanohelices of smaller diameter size (or with longer
wavelengths, for example, microwaves and radio waves). 

The macroscopic conducting coil driven by a household line voltage
alternating at $60\,\textup{Hz}$ generates electromagnetic waves
radiating at $60\,\textup{Hz}.$ Such radiation would have a wavelength
of roughly $5000\,\textup{km}$ in air, i.e., $\lambda=3\times10^{8}/60.$
A secondary conductive coil placed nearby it would perceive as if
the magnetic part of the radiating electromagnetic wave was a \textbf{DC}
magnetic field switching in an empty between on and off modes, except
now at the rate of $60\,\textup{Hz}$ cycle. Again, the argument holds
because even a solenoid of diameter as large as $100\,\textup{cm}$
is still fifty million times small compared to the wavelength of $\lambda=5000\,\textup{km}.$ 

In summary, the case involving nanohelix and the \textbf{IR} spectrum
of electromagnetic waves, in principle, is just the scaled down analogue
of the case which involves the interaction between electromagnetic
waves radiated by macroscopic conductive primary coil driven by a
household line voltage alternating at $60\,\textup{Hz}$ and the secondary
solenoid placed nearby. 

That being said, under the exposure of electromagnetic radiation,
net electric current gets induced inside the nanohelix in accordance
with Faraday's law of induction and the total induced current density
inside the nanohelix must be expressed as \[
\mathbf{J}=\sigma\mathbf{E}_{\textnormal{\textup{p}}}+\mathbf{J}_{\textup{B}},\]
 where $\mathbf{J}_{\textup{B}}$ is the induced current density contribution
arising from the magnetic part of the incidence ambient electromagnetic
wave. Nevertheless, for the analysis here, I neglect $\mathbf{J}_{\textup{B}}$
as this involves very lengthy derivation on its own. However, it is
reminded that $\mathbf{J}_{\textup{B}}$ only makes $\mathbf{J}$
bigger. Therefore, once I show that  Eq. (\ref{eq:amplification_proof})
is satisfied even with contribution from $\mathbf{J}_{\textup{B}}$
neglected in $\mathbf{J},$ redoing the problem with contribution
from $\mathbf{J}_{\textup{B}}$ included in $\mathbf{J}$ should only
make the case even firmer. 

\begin{figure}
\begin{centering}
\includegraphics[scale=0.4]{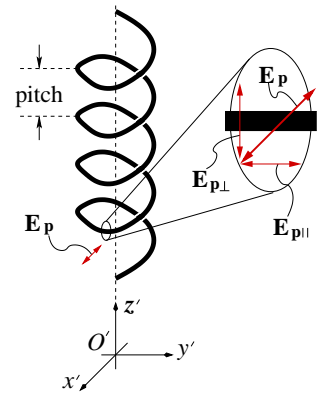}
\par\end{centering}

\caption{\label{Pol-of-PW}Perpendicular and parallel polarization components
at the local helix segment.}

\end{figure}

Returning from a short detour, in the case where finite helix is made
of a filamentary wire, only the component of electric field which
is parallel to the local length of wire can induce current as depicted
in Fig. \ref{Pol-of-PW}.  The polarization of impinging plane wave
can be decomposed into $\mathbf{E}_{\textnormal{\textup{p}}\perp}$
and $\mathbf{E}_{\textnormal{\textup{p}}\parallel}$ at the local
point $\mathbf{R}_{\textup{s}}$ of the finite helix, where $\mathbf{E}_{\textnormal{\textup{p}}\perp}$
and $\mathbf{E}_{\textnormal{\textup{p}}\parallel}$ are the two components
of $\mathbf{E}_{\textnormal{\textup{p}}}$ that are, respectively,
perpendicular and parallel to the local tangent of finite helix at
$\mathbf{R}_{\textup{s}}.$ In a filamentary wire, only $\mathbf{E}_{\textnormal{\textup{p}}\parallel}$
can result in induced current. Mathematically, $\mathbf{E}_{\textnormal{\textup{p}}\parallel}$
at local point, $\mathbf{R}_{\textup{s}},$ is expressed as \begin{equation}
\mathbf{E}_{\textnormal{\textup{p}}\parallel}=\left(\frac{\mathbf{E}_{\textnormal{\textup{p}}}\cdot\mathbf{T}}{\mathbf{T}\cdot\mathbf{T}}\right)\mathbf{T},\quad\left\{ \begin{array}{c}
\mathbf{E}_{\textnormal{\textup{p}}}\equiv\mathbf{E}_{\textnormal{\textup{p}}}\left(\mathbf{R}_{\textup{s}}\right),\\
\mathbf{T}\equiv\mathbf{T}\left(\mathbf{R}_{\textup{s}}\right),\end{array}\right.\label{eq:E-parallel-wire}\end{equation}
 where $\mathbf{T}$ is the local tangent vector for finite helix
at $\mathbf{R}_{\textup{s}}.$ The explicit expression for $\mathbf{T}$
has already been defined in Eq. (\ref{eq:T1-T2-T3-Def}). Insertion
of Eq. (\ref{eq:E-parallel-wire}) for $\mathbf{E}_{\textnormal{\textup{p}}}$
in Eq. (\ref{eq:Ohms-J}) ensures that only $\mathbf{E}_{\textnormal{\textup{p}}\parallel}$
takes part in the generation of locally induced current density, \begin{equation}
\mathbf{J}=\sigma\left(\frac{\mathbf{E}_{\textnormal{\textup{p}}}\cdot\mathbf{T}}{\mathbf{T}\cdot\mathbf{T}}\right)\mathbf{T},\quad\mathbf{J}\equiv\mathbf{J}\left(\mathbf{R}_{\textup{s}}\right).\label{eq:Ohms-J}\end{equation}

Without loss of generality, the electric field of impinging plane
wave at points on the finite helix can be expressed as \begin{equation}
\mathbf{E}_{\textnormal{\textup{p}}}\left(\mathbf{R}_{\textup{s}}\right)=\exp\left(i\mathbf{K}\cdot\mathbf{R}_{\textup{s}}+i\omega t\right)\sum_{i=1}^{3}E_{\textnormal{\textup{p}}i}\mathbf{e}_{i},\label{eq:PW-E-pre}\end{equation}
 where the wave vector $\mathbf{K}$ is given by \begin{equation}
\mathbf{K}=\sum_{i=1}^{3}K_{i}\mathbf{e}_{i}.\label{eq:K-vector}\end{equation}
 In terms of the direction cosines, \begin{equation}
\cos\alpha_{i}=\frac{E_{\textnormal{\textup{p}}i}}{E_{\textnormal{\textup{p}}}}\equiv\frac{E_{\textnormal{\textup{p}}i}}{\left\Vert \mathbf{E}_{\textnormal{\textup{p}}}\right\Vert },\quad i=1,2,3,\label{eq:direction-cosine}\end{equation}
 the Eq. (\ref{eq:PW-E-pre}) can be expressed as \begin{equation}
\mathbf{E}_{\textup{\textnormal{p}}}=E_{\textnormal{\textup{p}}}\exp\left(i\mathbf{K}\cdot\mathbf{R}_{\textup{s}}+i\omega t\right)\sum_{i=1}^{3}\cos\alpha_{i}\mathbf{e}_{i}.\label{eq:PW-E}\end{equation}
 With the following expressions, \begin{alignat*}{1}
\mathbf{T}\cdot\mathbf{T} & =a^{2}+b^{2},\\
\mathbf{E}_{\textnormal{\textup{p}}}\cdot\mathbf{T} & =E_{\textnormal{\textup{p}}}\chi\exp\left(i\mathbf{K}\cdot\mathbf{R}_{\textup{s}}+i\omega t\right),\end{alignat*}
 where \begin{equation}
\chi=-a\sin s\cos\alpha_{1}+a\cos s\cos\alpha_{2}+b\cos\alpha_{3},\label{eq:Xi-pre}\end{equation}
 the $\mathbf{J}$ of Eq. (\ref{eq:Ohms-J}) becomes \begin{equation}
\mathbf{J}^{\textnormal{\textup{TD}}}=\frac{\sigma E_{\textnormal{\textup{p}}}\chi}{a^{2}+b^{2}}\sum_{i=1}^{3}T_{i}\exp\left(i\mathbf{K}\cdot\mathbf{R}_{\textup{s}}+i\omega t\right)\mathbf{e}_{i}\label{eq:J-TD}\end{equation}
 with the superscript $\textnormal{\textup{TD}}$ denoting the time
domain. In the frequency domain analysis Eq. (\ref{eq:J-TD}) simplifies
to become \begin{equation}
\mathbf{J}\equiv\mathbf{J}^{\textnormal{\textup{FD}}}=\frac{\sigma E_{\textnormal{\textup{p}}}\chi}{a^{2}+b^{2}}\sum_{i=1}^{3}T_{i}\exp\left(i\mathbf{K}\cdot\mathbf{R}_{\textup{s}}\right)\mathbf{e}_{i},\label{eq:J-FD}\end{equation}
 where the superscript $\textup{\textnormal{FD}}$ now denotes the
frequency domain analysis, and $T_{i}$ and $\chi$ are respectively
from Eqs. (\ref{eq:T1-T2-T3-Def}) and (\ref{eq:Xi-pre}). 

It helps to simplify the analysis in the proceeding sections by re-expressing
Eq. (\ref{eq:Xi-pre}) in an alternate form. It is well known in mathematics
that any linear combination of sine waves of same period but different
phase shifts is also a sine wave of same period, but different phase
shift. It can be shown then \begin{align}
 & a_{1}\sin\varphi+a_{2}\sin\left(\varphi+b_{1}\right)\nonumber \\
 & =\sqrt{a_{1}^{2}+a_{2}^{2}+2a_{1}a_{2}\cos b_{1}}\nonumber \\
 & \times\sin\left[\varphi+\arctan\left(\frac{a_{2}\sin b_{1}}{a_{1}+a_{2}\cos b_{1}}\right)\right].\label{eq:sin-plus-sin}\end{align}
 Since \[
\sin\left(\varphi+b_{1}\right)=\cos\varphi,\quad\left\{ \begin{array}{c}
b_{1}=\pm\left(n+\frac{\pi}{2}\right),\\
n=0,1,2,\cdots,\end{array}\right.\]
 Eq. (\ref{eq:sin-plus-sin}) may be simplified to \begin{align}
a_{1}\sin\varphi+a_{2}\cos\varphi & =\sqrt{a_{1}^{2}+a_{2}^{2}}\nonumber \\
 & \times\sin\left[\varphi+\arctan\left(\frac{a_{2}}{a_{1}}\right)\right]\label{eq:sin-plus-cos}\end{align}
 for $b_{1}=\pi/2.$ Equation (\ref{eq:Xi-pre}) is compared with
Eq. (\ref{eq:sin-plus-cos}) to yield \begin{align}
\chi & =a\sqrt{\cos^{2}\alpha_{1}+\cos^{2}\alpha_{2}}\nonumber \\
 & \times\sin\left[s-\arctan\left(\frac{\cos\alpha_{2}}{\cos\alpha_{1}}\right)\right]+b\cos\alpha_{3},\label{eq:Xi}\end{align}
 where the odd property of arc tangent function, \[
\arctan\left(-\theta\right)=-\arctan\theta,\]
 has been utilized in the final step.

\subsection{Induced fields}

\subsubsection{Induced vector potential  }

In Cartesian coordinates, one has \[
u\equiv\left\Vert \mathbf{R}-\mathbf{R}_{\textup{s}}\right\Vert =\left[\sum_{i=1}^{3}\left(x_{i}-x_{\textup{o}i}-x_{i}^{\prime}\right)^{2}\right]^{1/2},\]
 where $\mathbf{R}=\sum_{i=1}^{3}x_{i}\mathbf{e}_{i}$ and $\mathbf{R}_{\textup{s}}$
is from Eq. (\ref{eq:Rs}). With Eq. (\ref{eq:x-y-z-prime}) substituted
in, $u$ becomes \begin{align}
u & =\left[\left(x_{1}-x_{\textup{o}1}-a\cos s\right)^{2}+\left(x_{2}-x_{\textup{o}2}-a\sin s\right)^{2}\right.\nonumber \\
 & \left.+\left(x_{3}-x_{\textup{o}3}-bs\right)^{2}\right]^{\frac{1}{2}}.\label{eq:u}\end{align}
  Insertion of Eqs. (\ref{eq:J-FD}) and (\ref{eq:u}) into Eq. (\ref{eq:A-integral-1D})
yields \begin{equation}
\mathbf{A}=\frac{\gamma\sigma E_{\textnormal{p}}}{a^{2}+b^{2}}\sum_{i=1}^{3}\mathbf{e}_{i}G_{i},\label{eq:A}\end{equation}
 where \begin{equation}
G_{i}=\int_{0}^{c}\frac{\chi T_{i}}{u}\exp\left(i\mathbf{K}\cdot\mathbf{R}_{\textup{s}}-iKu\right)ds.\label{eq:Gi-00}\end{equation}
 The quantity $\mathbf{K}\cdot\mathbf{R}_{\textup{s}}$ in Eq. (\ref{eq:Gi-00})
is given by \begin{equation}
\mathbf{K}\cdot\mathbf{R}_{\textup{s}}=\sum_{i=1}^{3}\left(K_{i}x_{\textup{o}i}+K_{i}x_{i}^{\prime}\right),\label{eq:K-dot-Rs-pre}\end{equation}
 where \begin{equation}
\sum_{i=1}^{3}K_{i}x_{i}^{\prime}=aK_{1}\cos s+aK_{2}\sin s+bK_{3}s.\label{eq:SUM-KiXi'}\end{equation}
 Equation (\ref{eq:SUM-KiXi'}) is compared with Eq. (\ref{eq:sin-plus-cos})
to yield \begin{align*}
\sum_{i=1}^{3}K_{i}x_{i}^{\prime} & =a\sqrt{K_{1}^{2}+K_{2}^{2}}\sin\left[s+\arctan\left(\frac{K_{1}}{K_{2}}\right)\right]\\
 & +bK_{3}s\end{align*}
 and Eq. (\ref{eq:K-dot-Rs-pre}) hence may be expressed \begin{align}
\mathbf{K}\cdot\mathbf{R}_{\textup{\textup{s}}} & =K_{1}x_{o1}+K_{2}x_{o2}+K_{3}\left(x_{o3}+bs\right)\nonumber \\
 & +a\sqrt{K_{1}^{2}+K_{2}^{2}}\sin\left[s+\arctan\left(\frac{K_{1}}{K_{2}}\right)\right].\label{eq:K-dot-Rs}\end{align}
 With Eq. (\ref{eq:K-dot-Rs}), the $G_{i}$ of Eq. (\ref{eq:Gi-00})
becomes \begin{equation}
G_{i}=\int_{0}^{c}\frac{\chi T_{i}}{u}\exp\left(iv\right)ds,\label{eq:Gi}\end{equation}
 where \begin{align}
v & =K_{1}x_{o1}+K_{2}x_{o2}+K_{3}\left(x_{o3}+bs\right)-Ku\nonumber \\
 & +a\sqrt{K_{1}^{2}+K_{2}^{2}}\sin\left[s+\arctan\left(\frac{K_{1}}{K_{2}}\right)\right].\label{eq:v}\end{align}
 Utilizing Euler's formula, \begin{equation}
\exp\left(iv\right)=\cos v+i\sin v,\label{eq:Eulers-formula}\end{equation}
 Eq. (\ref{eq:Gi}) may be separated into the real and imaginary parts\[
G_{i}=\Re G_{i}+i\Im G_{i},\]
 \begin{equation}
\left(\begin{array}{c}
\Re G_{i}\\
\Im G_{i}\end{array}\right)=\int_{0}^{c}\frac{\chi T_{i}}{u}\left(\begin{array}{c}
\cos v\\
\sin v\end{array}\right)ds,\label{eq:ReGi-and-ImGi}\end{equation}
 to yield\[
\mathbf{A}=\Re\mathbf{A}+i\Im\mathbf{A},\]
 \begin{equation}
\left(\begin{array}{c}
\Re\mathbf{A}\\
\Im\mathbf{A}\end{array}\right)=\frac{\gamma\sigma E_{\textnormal{p}}}{a^{2}+b^{2}}\sum_{i=1}^{3}\mathbf{e}_{i}\left(\begin{array}{c}
\Re G_{i}\\
\Im G_{i}\end{array}\right).\label{eq:ReA-and-ImA}\end{equation}
 In explicit form, the $\chi T_{i}$ in Eq. (\ref{eq:ReGi-and-ImGi}),
with the aid of Eqs. (\ref{eq:T1-T2-T3-Def}) and (\ref{eq:Xi}),
for each $i=1,$ $2,$ and $3,$ becomes \begin{alignat}{1}
\chi T_{1} & =-a^{2}c_{1}\sin\left(s-c_{2}\right)\sin s-abc_{3}\sin s,\label{eq:Chi-T1}\\
\chi T_{2} & =a^{2}c_{1}\sin\left(s-c_{2}\right)\cos s+abc_{3}\cos s,\label{eq:Chi-T2}\\
\chi T_{3} & =abc_{1}\sin\left(s-c_{2}\right)+b^{2}c_{3},\label{eq:Chi-T3}\end{alignat}
 where, \begin{align}
c_{1} & =\sqrt{\cos^{2}\alpha_{1}+\cos^{2}\alpha_{2}},\label{eq:c1-Def}\\
c_{2} & =\arctan\left(\frac{\cos\alpha_{2}}{\cos\alpha_{1}}\right),\label{eq:c2-Def}\\
c_{3} & =\cos\alpha_{3}.\label{eq:c3-Def}\end{align}
 Using trigonometric identities, \begin{alignat}{1}
\cos\theta\cos\varphi & =\frac{1}{2}\left[\cos\left(\theta-\varphi\right)+\cos\left(\theta+\varphi\right)\right],\label{eq:coscos-identity}\\
\sin\theta\sin\varphi & =\frac{1}{2}\left[\cos\left(\theta-\varphi\right)-\cos\left(\theta+\varphi\right)\right],\label{eq:sinsin-identity}\\
\sin\theta\cos\varphi & =\frac{1}{2}\left[\sin\left(\theta-\varphi\right)+\sin\left(\theta+\varphi\right)\right],\label{eq:sincos-identity}\end{alignat}
 the $\chi T_{1}$ of Eq. (\ref{eq:Chi-T1}) and $\chi T_{2}$ of
Eq. (\ref{eq:Chi-T2}) may be re-expressed into a canonical form,
\begin{alignat}{1}
\chi T_{1} & =\frac{1}{2}a^{2}c_{1}\left[\cos\left(2s-c_{2}\right)-\cos c_{2}\right]-abc_{3}\sin s,\label{eq:Chi-T1-canonical}\\
\chi T_{2} & =\frac{1}{2}a^{2}c_{1}\left[\sin\left(2s-c_{2}\right)-\sin c_{2}\right]+abc_{3}\cos s,\label{eq:Chi-T2-canonical}\end{alignat}
 where $c_{1},$ $c_{2},$ and $c_{3}$ are defined in Eqs. (\ref{eq:c1-Def})
thru (\ref{eq:c3-Def}) and the even and odd properties of cosine
and sine, \[
\cos\left(-x\right)=\cos x,\quad\sin\left(-x\right)=-\sin x,\]
 have been utilized in the result. The $\chi T_{3}$ of Eq. (\ref{eq:Chi-T1}),
$\chi T_{1}$ of Eq. (\ref{eq:Chi-T1-canonical}), and $\chi T_{2}$
of Eq. (\ref{eq:Chi-T2-canonical}) are substituted into Eq. (\ref{eq:ReGi-and-ImGi})
to yield  

\begin{align}
\Re G_{1} & =\frac{a}{4}\int_{0}^{c}\left[\vphantom{\frac{\frac{1}{1}}{1}}\frac{ac_{1}}{u}\cos\left(v+2s-c_{2}\right)-\frac{2ac_{1}\cos c_{2}}{u}\cos v\right.\nonumber \\
 & +\frac{ac_{1}}{u}\cos\left(v-2s+c_{2}\right)-\frac{2bc_{3}}{u}\sin\left(v+s\right)\nonumber \\
 & \left.+\frac{2bc_{3}}{u}\sin\left(v-s\right)\vphantom{\frac{\frac{1}{1}}{1}}\right]ds,\label{eq:ReG1}\end{align}
 \begin{align}
\Re G_{2} & =\frac{a}{4}\int_{0}^{c}\left[\vphantom{\frac{\frac{1}{1}}{1}}\frac{2bc_{3}}{u}\cos\left(v+s\right)-\frac{2ac_{1}}{u}\sin c_{2}\cos v\right.\nonumber \\
 & +\frac{2bc_{3}}{u}\cos\left(v-s\right)+\frac{ac_{1}}{u}\sin\left(v+2s-c_{2}\right)\nonumber \\
 & \left.-\frac{ac_{1}}{u}\sin\left(v-2s+c_{2}\right)\vphantom{\frac{\frac{1}{1}}{1}}\right]ds,\label{eq:ReG2}\end{align}
 \begin{align}
\Re G_{3} & =\frac{b}{2}\int_{0}^{c}\left[\vphantom{\frac{\frac{1}{1}}{1}}\frac{2bc_{3}}{u}\cos v+\frac{ac_{1}}{u}\sin\left(v+s-c_{2}\right)\right.\nonumber \\
 & \left.-\frac{ac_{1}}{u}\sin\left(v-s+c_{2}\right)\vphantom{\frac{\frac{1}{1}}{1}}\right]ds,\label{eq:ReG3}\end{align}
 \begin{align}
\Im G_{1} & =\frac{a}{4}\int_{0}^{c}\left[\vphantom{\frac{\frac{1}{1}}{1}}\frac{2bc_{3}}{u}\cos\left(v+s\right)-\frac{2bc_{3}}{u}\cos\left(v-s\right)\right.\nonumber \\
 & +\frac{ac_{1}}{u}\sin\left(v+2s-c_{2}\right)-\frac{2ac_{1}}{u}\cos c_{2}\sin v\nonumber \\
 & \left.+\frac{ac_{1}}{u}\sin\left(v-2s+c_{2}\right)\vphantom{\frac{\frac{1}{1}}{1}}\right]ds,\label{eq:ImG1}\end{align}
 \begin{align}
\Im G_{2} & =\frac{a}{4}\int_{0}^{c}\left[\vphantom{\frac{\frac{1}{1}}{1}}-\frac{ac_{1}}{u}\cos\left(v+2s-c_{2}\right)\right.\nonumber \\
 & +\frac{ac_{1}}{u}\cos\left(v-2s+c_{2}\right)+\frac{2bc_{3}}{u}\sin\left(v+s\right)\nonumber \\
 & \left.-\frac{2ac_{1}}{u}\sin c_{2}\sin v+\frac{2bc_{3}}{u}\sin\left(v-s\right)\vphantom{\frac{\frac{1}{1}}{1}}\right]ds,\label{eq:ImG2}\end{align}
 \begin{align}
\Im G_{3} & =\frac{b}{2}\int_{0}^{c}\left[\vphantom{\frac{\frac{1}{1}}{1}}-\frac{ac_{1}}{u}\cos\left(v+s-c_{2}\right)\right.\nonumber \\
 & \left.+\frac{ac_{1}}{u}\cos\left(v-s+c_{2}\right)+\frac{bc_{3}}{u}\sin v\vphantom{\frac{\frac{1}{1}}{1}}\right]ds,\label{eq:ImG3}\end{align}
 where $c_{1},$ $c_{2},$ and $c_{3}$ are defined in Eq. (\ref{eq:c1-Def}-\ref{eq:c3-Def}).

\subsubsection{Induced magnetic induction}

Substitution of $\mathbf{A}=\Re\mathbf{A}+i\Im\mathbf{A}$ into Eq.
(\ref{eq:curlA}) gives \[
\mathbf{B}=\nabla\times\Re\mathbf{A}+i\nabla\times\Im\mathbf{A}.\]
 Insertion of $\Re\mathbf{A}$ and $\Im\mathbf{A}$ from Eq. (\ref{eq:ReA-and-ImA})
yields \begin{equation}
\mathbf{B}=\Re\mathbf{B}+i\Im\mathbf{B},\label{eq:B-eq-RB-plus-IB}\end{equation}
 where \begin{equation}
\left(\begin{array}{c}
\Re\mathbf{B}\\
\Im\mathbf{B}\end{array}\right)=\frac{\gamma\sigma E_{\textnormal{p}}}{a^{2}+b^{2}}\underbrace{\nabla\times\sum_{i=1}^{3}\left(\begin{array}{c}
\Re G_{i}\\
\Im G_{i}\end{array}\right)\mathbf{e}_{i}}_{\textup{curl}}.\label{eq:ReB-ImB}\end{equation}
 The $i\textup{th}$ component of $\textup{curl }$ of Eq. (\ref{eq:ReB-ImB})
is  given by \[
\left[\nabla\times\sum_{i=1}^{3}\left(\begin{array}{c}
\Re G_{i}\\
\Im G_{i}\end{array}\right)\mathbf{e}_{i}\right]_{i}=\epsilon_{ijk}\partial_{j}\left(\begin{array}{c}
\Re G_{k}\\
\Im G_{k}\end{array}\right).\]
 In terms of vector components, Eq. (\ref{eq:ReB-ImB}) becomes \begin{equation}
\left(\begin{array}{c}
\Re\mathbf{B}\\
\Im\mathbf{B}\end{array}\right)=\frac{\gamma\sigma E_{\textnormal{p}}}{a^{2}+b^{2}}\sum_{i=1}^{3}\mathbf{e}_{i}\epsilon_{ijk}\partial_{j}\left(\begin{array}{c}
\Re G_{k}\\
\Im G_{k}\end{array}\right),\label{eq:ReB-ImB-1}\end{equation}
  where indices $j$ and $k$ are chosen in accordance with the cyclic
rule, \begin{equation}
\textnormal{If\,}i=\left(\begin{array}{c}
1\\
2\\
3\end{array}\right)\textnormal{\, then\,}j=\left(\begin{array}{c}
2\\
3\\
1\end{array}\right)\textnormal{ and }k=\left(\begin{array}{c}
3\\
1\\
2\end{array}\right),\label{eq:cyclic-rule}\end{equation}
 and $\epsilon_{ijk}$ is the Levi-Civita coefficient, \begin{equation}
\epsilon_{ijk}=\left\{ \begin{array}{ccc}
+1, &  & \textnormal{ if }\left(ijk\right)\textnormal{ is }\left(123\right)\textnormal{, }\left(231\right)\textnormal{, \textnormal{or} }\left(312\right)\\
-1, &  & \textnormal{ if }\left(ijk\right)\textnormal{ is }\left(321\right)\textnormal{, }\left(213\right)\textnormal{, \textnormal{or} }\left(132\right)\\
\;\:0, &  & \textnormal{\textup{otherwise}}\end{array}\right..\label{eq:Levi-Civita}\end{equation}
  Expanding out the Levi-Civita coefficient $\epsilon_{ijk},$ following
the rule stated in Eq. (\ref{eq:Levi-Civita}), the $\Re\mathbf{B}$
and $\Im\mathbf{B}$ of Eq. (\ref{eq:ReB-ImB-1}) become \begin{equation}
\left(\begin{array}{c}
\Re\mathbf{B}\\
\Im\mathbf{B}\end{array}\right)=\frac{\gamma\sigma E_{\textnormal{p}}}{a^{2}+b^{2}}\sum_{i=1}^{3}\left(\begin{array}{c}
\partial_{j}\Re G_{k}-\partial_{k}\Re G_{j}\\
\partial_{j}\Im G_{k}-\partial_{k}\Im G_{j}\end{array}\right)\mathbf{e}_{i},\label{eq:RB-IB-expanded}\end{equation}
 where the indices $j$ and $k$ are assigned in accordance with the
cyclic rule defined in Eq. (\ref{eq:cyclic-rule}). Equations (\ref{eq:ReG1})
thru (\ref{eq:ImG3}) may be summarized in the following form, \begin{equation}
\psi=\int\left[\sum_{\varrho}\frac{\zeta_{\textup{s}\varrho}\sin\left(v+\vartheta_{\textup{s}\varrho}\right)}{u}+\sum_{\iota}\frac{\zeta_{\textup{c}\iota}\cos\left(v+\vartheta_{\textup{c}\iota}\right)}{u}\right]ds,\label{eq:psi}\end{equation}
 where $\psi$ represents $\Re G_{i}$ or $\Im G_{i},$ the sums $\sum_{\varrho}$
and $\sum_{\iota}$ denote summation over terms involving sines and
cosines, respectively; and, $\left(\zeta_{\textup{s}\varrho},\zeta_{\textup{c}\iota}\right)$
and $\left(\vartheta_{\textup{s}\varrho},\vartheta_{\textup{c}\iota}\right)$
are the respective constant terms which can be identified from the
observation of sines and cosines of which involve  $v$ in their argument.
The operator $\nabla$ in Eq. (\ref{eq:ReB-ImB}) operates only on
the coordinates of the detector. Since only $u$ and $v$ involves
the detector coordinates, it can be shown that  \begin{equation}
\partial_{l}\psi=\int\left(\sum_{\varrho}\frac{\zeta_{\textup{s}\varrho}}{u}\psi_{\textup{s}\varrho}-\sum_{\iota}\frac{\zeta_{\textup{c}\iota}}{u}\psi_{\textup{c}\iota}\right)ds,\label{eq:par-l-psi}\end{equation}
  where $l=x,y,z$ or $l=1,2,3$ and \[
\psi_{\textup{s}\varrho}=\cos\left(v+\vartheta_{\textup{s}\varrho}\right)\partial_{l}v-\frac{\sin\left(v+\vartheta_{\textup{s}\varrho}\right)\partial_{l}u}{u},\]
 \[
\psi_{\textup{c}\iota}=\sin\left(v+\vartheta_{\textup{c}\iota}\right)\partial_{l}v+\frac{\cos\left(v+\vartheta_{\textup{c}\iota}\right)\partial_{l}u}{u}.\]
 To compute for $\partial_{l}u,$ I utilize the vector operator identity,
\begin{equation}
\nabla\left\Vert \mathbf{R}-\mathbf{R}_{\textup{s}}\right\Vert ^{n}=n\left\Vert \mathbf{R}-\mathbf{R}_{\textup{s}}\right\Vert ^{n-2}\left(\mathbf{R}-\mathbf{R}_{\textup{s}}\right).\label{eq:VOI-335}\end{equation}
 With Eq. (\ref{eq:VOI-335}), $\partial_{l}\left\Vert \mathbf{R}-\mathbf{R}_{\textup{s}}\right\Vert ^{n}$
becomes \[
\partial_{l}\left\Vert \mathbf{R}-\mathbf{R}_{\textup{s}}\right\Vert ^{n}=\mathbf{e}_{l}\cdot\nabla\left\Vert \mathbf{R}-\mathbf{R}_{\textup{s}}\right\Vert ^{n}=\frac{n\mathbf{e}_{l}\cdot\left(\mathbf{R}-\mathbf{R}_{\textup{s}}\right)}{\left\Vert \mathbf{R}-\mathbf{R}_{\textup{s}}\right\Vert ^{2-n}}\]
 or \begin{equation}
\partial_{l}\left\Vert \mathbf{R}-\mathbf{R}_{\textup{s}}\right\Vert ^{n}=\frac{n\left(x_{l}-x_{\textup{s}l}\right)}{\left\Vert \mathbf{R}-\mathbf{R}_{\textup{s}}\right\Vert ^{2-n}}.\label{eq:VOI-335-par-l}\end{equation}
 Since $u=\left\Vert \mathbf{R}-\mathbf{R}_{\textup{s}}\right\Vert ,$
I have  \begin{equation}
\partial_{l}u=\frac{x_{l}-x_{\textup{s}l}}{u}=-\frac{u_{l}}{u},\quad u_{l}=x_{\textup{s}l}-x_{l}.\label{eq:partial-of-u-of-l}\end{equation}
 For the $\partial_{l}v,$ using Eq. (\ref{eq:v}) for $v,$ one obtains
\begin{equation}
\partial_{l}v=-K\partial_{l}u=\frac{Ku_{l}}{u},\label{eq:partial-of-v-of-l}\end{equation}
 where Eq. (\ref{eq:partial-of-u-of-l}) has been substituted in for
$\partial_{l}u.$ Insertion of Eqs. (\ref{eq:partial-of-u-of-l})
and (\ref{eq:partial-of-v-of-l}) into Eq. (\ref{eq:par-l-psi}) yields
the expressions for $\partial_{l}\Re G_{i}$ and $\partial_{l}\Im G_{i},$
 \begin{equation}
\partial_{l}\psi=\int\left(\sum_{\varrho}\frac{\zeta_{\textup{s}\varrho}}{u}\psi_{\textup{s}\varrho o}+\sum_{\iota}\frac{\zeta_{\textup{c}\iota}}{u}\psi_{\textup{c}\iota o}\right)ds,\label{eq:par-l-psi-re}\end{equation}
 where \[
\psi_{\textup{s}\varrho o}=\frac{u_{l}K\cos\left(v+\vartheta_{\textup{s}\varrho}\right)}{u}+\frac{u_{l}\sin\left(v+\vartheta_{\textup{s}\varrho}\right)}{u^{2}},\]
 \[
\psi_{\textup{c}\iota o}=\frac{u_{l}\cos\left(v+\vartheta_{\textup{c}\iota}\right)}{u^{2}}-\frac{u_{l}K\sin\left(v+\vartheta_{\textup{c}\iota}\right)}{u}.\]
 Comparing Eq. (\ref{eq:par-l-psi-re}) with Eq. (\ref{eq:psi}),
I obtain the following transformation rule, \begin{equation}
\left\{ \begin{array}{c}
\sin\left(v+\vartheta_{\textup{s}\varrho}\right)\rightarrow\psi_{\textup{s}\varrho o},\\
\cos\left(v+\vartheta_{\textup{c}\iota}\right)\rightarrow\psi_{\textup{c}\iota o},\end{array}\right.\label{eq:TF-sin-and-cos}\end{equation}
 where $\vartheta_{\textup{s}\varrho}$ and $\vartheta_{\textup{c}\iota}$
are extracted from the argument of cosines and sines by direct comparison.
The computations of $\partial_{l}\Re G_{i}$ and $\partial_{l}\Im G_{i}$
are done by simple replacements of sines and cosines in Eqs. (\ref{eq:ReG1})
thru (\ref{eq:ImG3}) following the rule defined in Eq. (\ref{eq:TF-sin-and-cos}).
This yields the expressions  \begin{equation}
\left(\begin{array}{c}
\partial_{l}\Re G_{i}\\
\partial_{l}\Im G_{i}\end{array}\right)=u_{l}\left(\begin{array}{c}
\Re\Psi_{i}\\
\Im\Psi_{i}\end{array}\right),\label{eq:par-l-RGi}\end{equation}
 where 

\begin{align}
\Re\Psi_{1} & =\frac{a}{4}\int_{0}^{c}\left[\vphantom{\frac{\frac{1}{1}}{\frac{1}{1}}}\frac{ac_{1}}{u^{3}}\cos\left(v+2s-c_{2}\right)-\frac{2bc_{3}K}{u^{2}}\cos\left(v+s\right)\right.\nonumber \\
 & -\frac{2ac_{1}\cos c_{2}}{u^{3}}\cos v+\frac{2bc_{3}K}{u^{2}}\cos\left(v-s\right)\nonumber \\
 & +\frac{ac_{1}}{u^{3}}\cos\left(v-2s+c_{2}\right)-\frac{ac_{1}K}{u^{2}}\sin\left(v+2s-c_{2}\right)\nonumber \\
 & -\frac{2bc_{3}}{u^{2}}\left(1+\frac{1}{u}\right)\sin\left(v+s\right)+\frac{2ac_{1}K\cos c_{2}}{u^{2}}\sin v\nonumber \\
 & \left.+\frac{2bc_{3}}{u^{3}}\sin\left(v-s\right)-\frac{ac_{1}K}{u^{2}}\sin\left(v-2s+c_{2}\right)\vphantom{\frac{\frac{1}{1}}{\frac{1}{1}}}\right]ds,\label{eq:RePSI1}\end{align}
 \begin{align}
\Re\Psi_{2} & =\frac{a}{4}\int_{0}^{c}\left[\vphantom{\frac{\frac{1}{1}}{\frac{1}{1}}}\frac{ac_{1}K}{u^{2}}\cos\left(v+2s-c_{2}\right)+\frac{2bc_{3}}{u^{3}}\cos\left(v+s\right)\right.\nonumber \\
 & -\frac{2ac_{1}\sin c_{2}}{u^{3}}\cos v+\frac{2bc_{3}}{u^{3}}\cos\left(v-s\right)\nonumber \\
 & -\frac{ac_{1}K}{u^{2}}\cos\left(v-2s+c_{2}\right)+\frac{ac_{1}}{u^{3}}\sin\left(v+2s-c_{2}\right)\nonumber \\
 & -\frac{2bc_{3}K}{u^{2}}\sin\left(v+s\right)+\frac{2ac_{1}K\sin c_{2}}{u^{2}}\sin v\nonumber \\
 & \left.-\frac{2bc_{3}K}{u^{2}}\sin\left(v-s\right)-\frac{ac_{1}}{u^{3}}\sin\left(v-2s+c_{2}\right)\vphantom{\frac{\frac{1}{1}}{\frac{1}{1}}}\right]ds,\label{eq:RePSI2}\end{align}
 \begin{align}
\Re\Psi_{3} & =\frac{b}{2}\int_{0}^{c}\left[\vphantom{\frac{\frac{1}{1}}{\frac{1}{1}}}\frac{ac_{1}K}{u^{2}}\cos\left(v+s-c_{2}\right)+\frac{2bc_{3}}{u^{3}}\cos v\right.\nonumber \\
 & -\frac{ac_{1}K}{u^{2}}\cos\left(v-s+c_{2}\right)+\frac{ac_{1}}{u^{3}}\sin\left(v+s-c_{2}\right)\nonumber \\
 & \left.-\frac{2bc_{3}K}{u^{2}}\sin v-\frac{ac_{1}}{u^{3}}\sin\left(v-s+c_{2}\right)\vphantom{\frac{\frac{1}{1}}{\frac{1}{1}}}\right]ds,\label{eq:RePSI3}\end{align}
 \begin{align}
\Im\Psi_{1} & =\frac{a}{4}\int_{0}^{c}\left[\vphantom{\frac{\frac{1}{1}}{\frac{1}{1}}}\frac{ac_{1}K}{u^{2}}\cos\left(v+2s-c_{2}\right)+\frac{2bc_{3}}{u^{3}}\cos\left(v+s\right)\right.\nonumber \\
 & -\frac{2ac_{1}K\cos c_{2}}{u^{2}}\cos v-\frac{2bc_{3}}{u^{3}}\cos\left(v-s\right)\nonumber \\
 & +\frac{ac_{1}K}{u^{2}}\cos\left(v-2s+c_{2}\right)+\frac{ac_{1}}{u^{3}}\sin\left(v+2s-c_{2}\right)\nonumber \\
 & -\frac{2bc_{3}K}{u^{2}}\sin\left(v+s\right)-\frac{2ac_{1}\cos c_{2}}{u^{3}}\sin v\nonumber \\
 & \left.+\frac{2bc_{3}K}{u^{2}}\sin\left(v-s\right)+\frac{ac_{1}}{u^{3}}\sin\left(v-2s+c_{2}\right)\vphantom{\frac{\frac{1}{1}}{\frac{1}{1}}}\right]ds,\label{eq:ImPSI1}\end{align}
 \begin{align}
\Im\Psi_{2} & =\frac{a}{4}\int_{0}^{c}\left[\vphantom{\frac{\frac{1}{1}}{\frac{1}{1}}}-\frac{ac_{1}}{u^{3}}\cos\left(v+2s-c_{2}\right)\right.\nonumber \\
 & +\frac{2bc_{3}K}{u^{2}}\cos\left(v+s\right)-\frac{2ac_{1}K\sin c_{2}}{u^{2}}\cos v\nonumber \\
 & +\frac{2bc_{3}K}{u^{2}}\cos\left(v-s\right)+\frac{ac_{1}}{u^{3}}\cos\left(v-2s+c_{2}\right)\nonumber \\
 & +\frac{ac_{1}K}{u^{2}}\sin\left(v+2s-c_{2}\right)+\frac{2bc_{3}}{u^{3}}\sin\left(v+s\right)\nonumber \\
 & -\frac{2ac_{1}\sin c_{2}}{u^{3}}\sin v+\frac{2bc_{3}}{u^{3}}\sin\left(v-s\right)\nonumber \\
 & \left.-\frac{ac_{1}K}{u^{2}}\sin\left(v-2s+c_{2}\right)\vphantom{\frac{\frac{1}{1}}{\frac{1}{1}}}\right]ds,\label{eq:ImPSI2}\end{align}
 \begin{align}
\Im\Psi_{3} & =\frac{b}{2}\int_{0}^{c}\left[\vphantom{\frac{\frac{1}{1}}{\frac{1}{1}}}-\frac{ac_{1}}{u^{3}}\cos\left(v+s-c_{2}\right)+\frac{bc_{3}K}{u^{2}}\cos v\right.\nonumber \\
 & +\frac{ac_{1}}{u^{3}}\cos\left(v-s+c_{2}\right)+\frac{ac_{1}K}{u^{2}}\sin\left(v+s-c_{2}\right)\nonumber \\
 & \left.+\frac{bc_{3}}{u^{3}}\sin v-\frac{ac_{1}K}{u^{2}}\sin\left(v-s+c_{2}\right)\vphantom{\frac{\frac{1}{1}}{\frac{1}{1}}}\right]ds,\label{eq:ImPSI3}\end{align}
 where $c_{1},$ $c_{2},$ and $c_{3}$ are defined in Eqs. (\ref{eq:c1-Def})
thru (\ref{eq:c3-Def}). 

The components of $\Re\mathbf{B}$ and $\Im\mathbf{B}$ are readily
extracted from Eq. (\ref{eq:RB-IB-expanded}) to yield \[
\left(\begin{array}{c}
\Re\mathbf{B}\\
\Im\mathbf{B}\end{array}\right)=\sum_{i=1}^{3}\left(\begin{array}{c}
\Re B_{i}\\
\Im B_{i}\end{array}\right)\mathbf{e}_{i},\]
  where \begin{equation}
\left(\begin{array}{c}
\Re B_{i}\\
\Im B_{i}\end{array}\right)=\frac{\gamma\sigma E_{\textnormal{p}}}{a^{2}+b^{2}}\left(\begin{array}{c}
\partial_{j}\Re G_{k}-\partial_{k}\Re G_{j}\\
\partial_{j}\Im G_{k}-\partial_{k}\Im G_{j}\end{array}\right).\label{eq:RBi-IBi-1}\end{equation}
 Utilizing Eq. (\ref{eq:par-l-RGi}), the $\Re B_{i}$ and $\Im B_{i}$
of Eq. (\ref{eq:RBi-IBi-1}) become \begin{equation}
\left(\begin{array}{c}
\Re B_{i}\\
\Im B_{i}\end{array}\right)=\frac{\gamma\sigma E_{\textnormal{p}}}{a^{2}+b^{2}}\left[u_{j}\left(\begin{array}{c}
\Re\Psi_{k}\\
\Im\Psi_{k}\end{array}\right)-u_{k}\left(\begin{array}{c}
\Re\Psi_{j}\\
\Im\Psi_{j}\end{array}\right)\right],\label{eq:RBi-IBi}\end{equation}
 where $i=1,2,3$ and the indices $j$ and $k$ are assigned in accordance
with the cyclic rule defined in Eq. (\ref{eq:cyclic-rule}). Knowing
that $B_{\textup{P}}=E_{\textnormal{p}}/c_{\textup{o}},$ where $c_{o}$
is the speed of light in vacuum, Eq. (\ref{eq:RBi-IBi}) may be expressed
as \begin{equation}
\left(\begin{array}{c}
\Re B_{i}\\
\Im B_{i}\end{array}\right)=\frac{\gamma\sigma c_{\textup{o}}B_{\textup{P}}}{a^{2}+b^{2}}\left[u_{j}\left(\begin{array}{c}
\Re\Psi_{k}\\
\Im\Psi_{k}\end{array}\right)-u_{k}\left(\begin{array}{c}
\Re\Psi_{j}\\
\Im\Psi_{j}\end{array}\right)\right].\label{eq:RBi-IBi-in-Bp}\end{equation}
 Since $\mathbf{B}=\Re\mathbf{B}+i\Im\mathbf{B}$ from Eq. (\ref{eq:B-eq-RB-plus-IB}),
the magnitude of $\mathbf{B}$ is given by \begin{alignat*}{1}
B & =\sqrt{\mathbf{B}^{\dagger}\mathbf{B}}=\sqrt{\left(\Re\mathbf{B}-i\Im\mathbf{B}\right)\cdot\left(\Re\mathbf{B}+i\Im\mathbf{B}\right)}\\
 & =\sqrt{\Re\mathbf{B}\cdot\Re\mathbf{B}+\Im\mathbf{B}\cdot\Im\mathbf{B}}\\
 & =\left\{ \sum_{i=1}^{3}\left[\left(\Re B_{i}\right)^{2}+\left(\Im B_{i}\right)^{2}\right]\right\} ^{1/2}.\end{alignat*}
 With Eq. (\ref{eq:RBi-IBi-in-Bp}) substituted in for $\Re B_{i}$
and $\Im B_{i},$ one obtains \begin{align}
\frac{B_{\textup{rad}}}{B_{\textup{P}}}\equiv\frac{B}{B_{\textup{P}}} & =\frac{\gamma\sigma c_{\textup{o}}}{a^{2}+b^{2}}\left[\left(u_{2}\Re\Psi_{3}-u_{3}\Re\Psi_{2}\right)^{2}\right.\nonumber \\
 & +\left(u_{3}\Re\Psi_{1}-u_{1}\Re\Psi_{3}\right)^{2}+\left(u_{1}\Re\Psi_{2}-u_{2}\Re\Psi_{1}\right)^{2}\nonumber \\
 & +\left(u_{2}\Im\Psi_{3}-u_{3}\Im\Psi_{2}\right)^{2}+\left(u_{3}\Im\Psi_{1}-u_{1}\Im\Psi_{3}\right)^{2}\nonumber \\
 & \left.+\left(u_{1}\Im\Psi_{2}-u_{2}\Im\Psi_{1}\right)^{2}\right]^{\frac{1}{2}},\label{eq:Brad-over-Bp}\end{align}
 where $\Re\Psi_{i}$ and $\Im\Psi_{i}$ are from Eqs. (\ref{eq:RePSI1})
thru (\ref{eq:ImPSI3}) and $u_{i}$ is defined in Eq. (\ref{eq:partial-of-u-of-l})
for each $i=1,2,3.$

\subsubsection{Induced electric field}

The associated electric field may be obtained from Eq. (\ref{eq:E-in-A}).
Insertion of Eq. (\ref{eq:A}) into Eq. (\ref{eq:E-in-A}) gives \begin{equation}
\mathbf{E}=-i\omega\eta\mathbf{A}-\frac{ig\sigma E_{\textnormal{p}}}{\omega\left(a^{2}+b^{2}\right)}\nabla\left(\nabla\cdot\sum_{l=1}^{3}\mathbf{e}_{l}G_{l}\right).\label{eq:E-expression-2}\end{equation}
 The term involving Cartesian gradient operator $\nabla,$ where $\nabla\equiv\sum_{i=1}^{3}\mathbf{e}_{i}\partial_{i},$
can be expressed as \begin{alignat*}{1}
\nabla\left(\nabla\cdot\sum_{l=1}^{3}\mathbf{e}_{l}G_{l}\right) & =\nabla\sum_{l=1}^{3}\partial_{l}G_{l}=\sum_{i=1}^{3}\mathbf{e}_{i}\partial_{i}\sum_{l=1}^{3}\partial_{l}G_{l}\\
 & =\sum_{i=1}^{3}\sum_{l=1}^{3}\mathbf{e}_{i}\partial_{i}\partial_{l}G_{l}\end{alignat*}
 and Eq. (\ref{eq:E-expression-2}) becomes \begin{equation}
\mathbf{E}=-i\omega\eta\mathbf{A}-\frac{ig\sigma E_{\textnormal{p}}}{\omega\left(a^{2}+b^{2}\right)}\sum_{i=1}^{3}\sum_{l=1}^{3}\mathbf{e}_{i}\partial_{i}\partial_{l}G_{l},\label{eq:E-expression-3}\end{equation}
  where $\partial_{i}\partial_{l}G_{l}\equiv\partial_{l}\partial_{i}G_{l}$
since $\partial^{2}/\left(\partial x\partial y\right)=\partial^{2}/\left(\partial y\partial x\right)$
for any mixed partial derivatives (recall that notation $\partial_{i}$
or $\partial_{x}$ represents $\partial/\partial x$). Insertion of
Eq. (\ref{eq:ReA-and-ImA}) into Eq. (\ref{eq:E-expression-3}) finally
yields \[
\mathbf{E}=\Re\mathbf{E}+i\Im\mathbf{E},\]
 where \begin{equation}
\Re\mathbf{E}=\omega\eta\Im\mathbf{A}+\frac{g\sigma E_{\textnormal{p}}}{\omega\left(a^{2}+b^{2}\right)}\sum_{i=1}^{3}\sum_{l=1}^{3}\mathbf{e}_{i}\partial_{i}\partial_{l}\Im G_{l},\label{eq:RE}\end{equation}
 \begin{equation}
\Im\mathbf{E}=-\omega\eta\Re\mathbf{A}-\frac{g\sigma E_{\textnormal{p}}}{\omega\left(a^{2}+b^{2}\right)}\sum_{i=1}^{3}\sum_{l=1}^{3}\mathbf{e}_{i}\partial_{i}\partial_{l}\Re G_{l}.\label{eq:IE}\end{equation}
 Utilizing Eq. (\ref{eq:par-l-RGi}) for $\partial_{l}\Im G_{l}$
and $\partial_{l}\Re G_{l},$ one has \[
\partial_{i}\partial_{l}\left(\begin{array}{c}
\Im G_{l}\\
\Re G_{l}\end{array}\right)=\partial_{i}\left[x_{\textup{s}l}\left(\begin{array}{c}
\Im\Psi_{l}\\
\Re\Psi_{l}\end{array}\right)-x_{l}\left(\begin{array}{c}
\Im\Psi_{l}\\
\Re\Psi_{l}\end{array}\right)\right]\]
 or \[
\partial_{i}\partial_{l}\left(\begin{array}{c}
\Im G_{l}\\
\Re G_{l}\end{array}\right)=u_{l}\partial_{i}\left(\begin{array}{c}
\Im\Psi_{l}\\
\Re\Psi_{l}\end{array}\right)-\left(\begin{array}{c}
\Im\Psi_{l}\\
\Re\Psi_{l}\end{array}\right)\partial_{i}x_{l}.\]
 Since the notation $\partial_{i}$ denotes $\partial/\partial x$
for, say, $i=1,$ and notation $x_{l}$ denotes $x$ for $l=1$, $y$
for $l=2$ and so on, one finds \begin{equation}
\partial_{i}x_{l}=\delta_{il},\quad\delta_{il}=\left\{ \begin{array}{ccc}
1, &  & \textnormal{if }i=l\\
0, &  & \textnormal{otherwise}\end{array}\right..\label{eq:Kronecker}\end{equation}
 where $\delta_{il}$ is the Kronecker delta. The partial derivatives
$\partial_{i}\partial_{l}\Im G_{l}$ and $\partial_{i}\partial_{l}\Re G_{l}$
hence become \[
\partial_{i}\partial_{l}\left(\begin{array}{c}
\Im G_{l}\\
\Re G_{l}\end{array}\right)=u_{l}\partial_{i}\left(\begin{array}{c}
\Im\Psi_{l}\\
\Re\Psi_{l}\end{array}\right)-\delta_{il}\left(\begin{array}{c}
\Im\Psi_{l}\\
\Re\Psi_{l}\end{array}\right)\]
 and the $\Re\mathbf{E}$ of Eq. (\ref{eq:RE}) and $\Im\mathbf{E}$
of Eq. (\ref{eq:IE}) get re-expressed as \begin{align}
\Re\mathbf{E} & =\frac{\sigma E_{\textnormal{p}}}{a^{2}+b^{2}}\sum_{i=1}^{3}\left[\vphantom{\frac{\frac{1}{1}}{\frac{1}{1}}}\gamma\omega\eta\Im G_{i}\right.\nonumber \\
 & \left.+\frac{g}{\omega}\sum_{l=1}^{3}\left(u_{l}\partial_{i}\Im\Psi_{l}-\delta_{il}\Im\Psi_{l}\right)\vphantom{\frac{\frac{1}{1}}{\frac{1}{1}}}\right]\mathbf{e}_{i},\label{eq:RE-gen}\end{align}
 \begin{align}
\Im\mathbf{E} & =-\frac{\sigma E_{\textnormal{p}}}{a^{2}+b^{2}}\sum_{i=1}^{3}\left[\vphantom{\frac{\frac{1}{1}}{\frac{1}{1}}}\gamma\omega\eta\Re G_{i}\right.\nonumber \\
 & \left.+\frac{g}{\omega}\sum_{l=1}^{3}\left(u_{l}\partial_{i}\Re\Psi_{l}-\delta_{il}\Re\Psi_{l}\right)\vphantom{\frac{\frac{1}{1}}{\frac{1}{1}}}\right]\mathbf{e}_{i},\label{eq:IE-gen}\end{align}
 where Eq. (\ref{eq:ReA-and-ImA}) has been substituted in for $\Re\mathbf{A}$
and $\Im\mathbf{A}.$ To compute for $\partial_{i}\Re\Psi_{l}$ and
$\partial_{i}\Im\Psi_{l},$ one notes that $\Re\Psi_{l}$ and $\Im\Psi_{l}$
can be summarized in form as \begin{align}
\Upsilon & =\int\left[\sum_{\varrho}\frac{\zeta_{\textup{s}\varrho}^{\textup{a}}}{u^{3}}\sin\left(v+\vartheta_{\textup{s}\varrho}^{\textup{a}}\right)\right.\nonumber \\
 & +\sum_{\sigma}\frac{\zeta_{\textup{s}\sigma}^{\textup{b}}}{u^{2}}\sin\left(v+\vartheta_{\textup{s}\sigma}^{\textup{b}}\right)+\sum_{\iota}\frac{\zeta_{\textup{c}\iota}^{\textup{c}}}{u^{3}}\cos\left(v+\vartheta_{\textup{c}\iota}^{\textup{c}}\right)\nonumber \\
 & \left.+\sum_{\varsigma}\frac{\zeta_{\textup{c}\varsigma}^{\textup{d}}}{u^{2}}\cos\left(v+\vartheta_{\textup{c}\varsigma}^{\textup{d}}\right)\right]ds,\label{eq:PSI-form}\end{align}
  where $\Upsilon$ represents $\Re\Psi_{l}$ or $\Im\Psi_{l},$ the
sums $\sum_{\varrho},$ $\sum_{\sigma},$ $\sum_{\iota},$ and $\sum_{\varsigma}$
denote summation over terms involving sines and cosines divided by
$u^{2}$ or $u^{3};$ and, $\left(\zeta_{\textup{s}\varrho}^{\textup{a}},\zeta_{\textup{s}\sigma}^{\textup{b}},\zeta_{\textup{c}\iota}^{\textup{c}},\zeta_{\textup{c}\varsigma}^{\textup{d}}\right)$
and $\left(\vartheta_{\textup{s}\varrho}^{\textup{a}},\vartheta_{\textup{s}\sigma}^{\textup{b}},\vartheta_{\textup{c}\iota}^{\textup{c}},\vartheta_{\textup{c}\varsigma}^{\textup{d}}\right)$
are the respective constant terms which can be identified from the
observation of sines and cosines of  which involve $v$ in their argument.
The operator $\partial_{i}$ only acts on non-source coordinates,
of course. Since $u$ and $v$ are the only terms with non-source
coordinates, one has \begin{align}
\partial_{i}\Upsilon & =\int\left(\sum_{\varrho}\zeta_{\textup{s}\varrho}^{\textup{a}}\Upsilon_{\varrho}^{\textup{a}}+\sum_{\sigma}\zeta_{\textup{s}\sigma}^{\textup{b}}\Upsilon_{\sigma}^{\textup{b}}\right.\nonumber \\
 & \left.+\sum_{\iota}\zeta_{\textup{c}\iota}^{\textup{c}}\Upsilon_{\iota}^{\textup{c}}+\sum_{\varsigma}\zeta_{\textup{c}\varsigma}^{\textup{d}}\Upsilon_{\varsigma}^{\textup{d}}\right)ds,\label{eq:par-of-PSI}\end{align}
 where $i=x,y,z$ or $i=1,2,3$ and \[
\Upsilon_{\varrho}^{\textup{a}}=\frac{\cos\left(v+\vartheta_{\textup{s}\varrho}^{\textup{a}}\right)\partial_{i}v}{u^{3}}-\frac{\sin\left(v+\vartheta_{\textup{s}\varrho}^{\textup{a}}\right)\partial_{i}u^{3}}{u^{6}},\]
 \[
\Upsilon_{\sigma}^{\textup{b}}=\frac{\cos\left(v+\vartheta_{\textup{s}\sigma}^{\textup{b}}\right)\partial_{i}v}{u^{2}}-\frac{\sin\left(v+\vartheta_{\textup{s}\sigma}^{\textup{b}}\right)\partial_{i}u^{2}}{u^{4}},\]
 \[
\Upsilon_{\iota}^{\textup{c}}=-\frac{\sin\left(v+\vartheta_{\textup{c}\iota}^{\textup{c}}\right)\partial_{i}v}{u^{3}}-\frac{\cos\left(v+\vartheta_{\textup{c}\iota}^{\textup{c}}\right)\partial_{i}u^{3}}{u^{6}},\]
 \[
\Upsilon_{\varsigma}^{\textup{d}}=-\frac{\sin\left(v+\vartheta_{\textup{c}\varsigma}^{\textup{d}}\right)\partial_{i}v}{u^{2}}-\frac{\cos\left(v+\vartheta_{\textup{c}\varsigma}^{\textup{d}}\right)\partial_{i}u^{2}}{u^{4}}.\]
  Utilizing Eq. (\ref{eq:VOI-335-par-l}), it can be shown \[
\partial_{i}u^{2}=2\left(x_{i}-x_{\textup{s}i}\right),\quad\partial_{i}u^{3}=3\left(x_{i}-x_{\textup{s}i}\right)u\]
 or, since $x_{i}-x_{\textup{s}i}=-u_{i},$ \begin{equation}
\partial_{i}u^{2}=-2u_{i},\quad\partial_{i}u^{3}=-3u_{i}u,\label{eq:par-u2-u3}\end{equation}
 where the dummy index $l$ has been replaced by another dummy index
$i,$ of course. The expression for $\partial_{i}v$ has already been
defined in Eq. (\ref{eq:partial-of-v-of-l}), i.e., let $i=l.$ With
Eqs. (\ref{eq:partial-of-v-of-l}) and (\ref{eq:par-u2-u3}), the
expression for $\partial_{i}\Upsilon$ of Eq. (\ref{eq:par-of-PSI})
becomes \begin{align}
\partial_{i}\Upsilon & =\int\left(\sum_{\varrho}\frac{\zeta_{\textup{s}\varrho}^{\textup{a}}}{u^{3}}\Upsilon_{\varrho o}^{\textup{a}}+\sum_{\sigma}\frac{\zeta_{\textup{s}\sigma}^{\textup{b}}}{u^{2}}\Upsilon_{\sigma o}^{\textup{b}}\right.\nonumber \\
 & \left.+\sum_{\iota}\frac{\zeta_{\textup{c}\iota}^{\textup{c}}}{u^{3}}\Upsilon_{\iota o}^{\textup{c}}+\sum_{\varsigma}\frac{\zeta_{\textup{c}\varsigma}^{\textup{d}}}{u^{2}}\Upsilon_{\varsigma o}^{\textup{d}}\right)ds,\label{eq:par-PSI-1}\end{align}
 where \[
\Upsilon_{\varrho o}^{\textup{a}}=\frac{u_{i}K\cos\left(v+\vartheta_{\textup{s}\varrho}^{\textup{a}}\right)}{u}+\frac{3u_{i}\sin\left(v+\vartheta_{\textup{s}\varrho}^{\textup{a}}\right)}{u^{2}},\]
 \[
\Upsilon_{\sigma o}^{\textup{b}}=\frac{u_{i}K\cos\left(v+\vartheta_{\textup{s}\sigma}^{\textup{b}}\right)}{u}+\frac{2u_{i}\sin\left(v+\vartheta_{\textup{s}\sigma}^{\textup{b}}\right)}{u^{2}},\]
 \[
\Upsilon_{\iota o}^{\textup{c}}=\frac{3u_{i}\cos\left(v+\vartheta_{\textup{c}\iota}^{\textup{c}}\right)}{u^{2}}-\frac{u_{i}K\sin\left(v+\vartheta_{\textup{c}\iota}^{\textup{c}}\right)}{u},\]
 \[
\Upsilon_{\varsigma o}^{\textup{d}}=\frac{2u_{i}\cos\left(v+\vartheta_{\textup{c}\varsigma}^{\textup{d}}\right)}{u^{2}}-\frac{u_{i}K\sin\left(v+\vartheta_{\textup{c}\varsigma}^{\textup{d}}\right)}{u}.\]
  Comparing Eq. (\ref{eq:par-PSI-1}) with Eq. (\ref{eq:PSI-form}),
one identifies the transformation rule for the sines and cosines given
by \begin{equation}
\left\{ \begin{array}{ccc}
\sin\left(v+\vartheta_{\textup{s}\varrho}^{\textup{a}}\right)\rightarrow\Upsilon_{\varrho o}^{\textup{a}}, &  & \sin\left(v+\vartheta_{\textup{s}\sigma}^{\textup{b}}\right)\rightarrow\Upsilon_{\sigma o}^{\textup{b}},\\
\cos\left(v+\vartheta_{\textup{c}\iota}^{\textup{c}}\right)\rightarrow\Upsilon_{\iota o}^{\textup{c}}, &  & \cos\left(v+\vartheta_{\textup{c}\varsigma}^{\textup{d}}\right)\rightarrow\Upsilon_{\varsigma o}^{\textup{d}},\end{array}\right.\label{eq:tran-E1-E2-E3-E4}\end{equation}
  where $\vartheta_{\textup{s}\varrho}^{\textup{a}},$ $\vartheta_{\textup{s}\sigma}^{\textup{b}},$
$\vartheta_{\textup{c}\iota}^{\textup{c}},$ and $\vartheta_{\textup{c}\varsigma}^{\textup{d}}$
can be identified by observing appropriate cosines or sines in expressions
for $\Re\Psi_{l}$ and $\Im\Psi_{l}$ of Eqs. (\ref{eq:RePSI1}) thru
(\ref{eq:ImPSI3}). Application of Eq. (\ref{eq:tran-E1-E2-E3-E4})
on Eqs. (\ref{eq:RePSI1}) thru (\ref{eq:ImPSI3}) yields  \begin{equation}
\left(\begin{array}{c}
\partial_{i}\Re\Psi_{l}\\
\partial_{i}\Im\Psi_{l}\end{array}\right)=u_{i}\left(\begin{array}{c}
\Re\Lambda_{l}\\
\Im\Lambda_{l}\end{array}\right),\label{eq:par-RPSI-IPSI}\end{equation}
 where 

\begin{align}
\Re\Lambda_{1} & =\frac{a}{4}\int_{0}^{c}\left[\vphantom{\frac{\frac{1}{1}}{\frac{1}{1}}}-ac_{1}\left(\frac{K^{2}}{u^{3}}-\frac{3}{u^{5}}\right)\cos\left(v+2s-c_{2}\right)\right.\nonumber \\
 & -2bc_{3}K\left(\frac{1}{u^{3}}+\frac{3}{u^{4}}\right)\cos\left(v+s\right)\nonumber \\
 & +2ac_{1}\cos c_{2}\left(\frac{K^{2}}{u^{3}}-\frac{3}{u^{5}}\right)\cos v\nonumber \\
 & -ac_{1}\left(\frac{K^{2}}{u^{3}}-\frac{3}{u^{5}}\right)\cos\left(v-2s+c_{2}\right)\nonumber \\
 & +\frac{6bc_{3}K}{u^{4}}\cos\left(v-s\right)-\frac{3ac_{1}K}{u^{4}}\sin\left(v+2s-c_{2}\right)\nonumber \\
 & +2bc_{3}\left(\frac{K^{2}}{u^{3}}-\frac{2}{u^{4}}-\frac{3}{u^{5}}\right)\sin\left(v+s\right)\nonumber \\
 & +\frac{6ac_{1}K\cos c_{2}}{u^{4}}\sin v-2bc_{3}\left(\frac{K^{2}}{u^{3}}-\frac{3}{u^{5}}\right)\sin\left(v-s\right)\nonumber \\
 & \left.-\frac{3ac_{1}K}{u^{4}}\sin\left(v-2s+c_{2}\right)\vphantom{\frac{\frac{1}{1}}{\frac{1}{1}}}\right]ds,\label{eq:ReLAMBDA1}\end{align}
 \begin{align}
\Re\Lambda_{2} & =\frac{a}{4}\int_{0}^{c}\left[\vphantom{\frac{\frac{1}{1}}{\frac{1}{1}}}\frac{3ac_{1}K}{u^{4}}\cos\left(v+2s-c_{2}\right)\right.\nonumber \\
 & -2bc_{3}\left(\frac{K^{2}}{u^{3}}-\frac{3}{u^{5}}\right)\cos\left(v+s\right)\nonumber \\
 & +2ac_{1}\sin c_{2}\left(\frac{K^{2}}{u^{3}}-\frac{3}{u^{5}}\right)\cos v\nonumber \\
 & -2bc_{3}\left(\frac{K^{2}}{u^{3}}-\frac{3}{u^{5}}\right)\cos\left(v-s\right)\nonumber \\
 & -\frac{3ac_{1}K}{u^{4}}\cos\left(v-2s+c_{2}\right)-\frac{6bc_{3}K}{u^{4}}\sin\left(v+s\right)\nonumber \\
 & -ac_{1}\left(\frac{K^{2}}{u^{3}}-\frac{3}{u^{5}}\right)\sin\left(v+2s-c_{2}\right)\nonumber \\
 & +\frac{6ac_{1}K\sin c_{2}}{u^{4}}\sin v-\frac{6bc_{3}K}{u^{4}}\sin\left(v-s\right)\nonumber \\
 & \left.+ac_{1}\left(\frac{K^{2}}{u^{3}}-\frac{3}{u^{5}}\right)\sin\left(v-2s+c_{2}\right)\vphantom{\frac{\frac{1}{1}}{\frac{1}{1}}}\right]ds,\label{eq:ReLAMBDA2}\end{align}
 \begin{align}
\Re\Lambda_{3} & =\frac{b}{2}\int_{0}^{c}\left[\vphantom{\frac{\frac{1}{1}}{\frac{1}{1}}}\frac{3ac_{1}K}{u^{4}}\cos\left(v+s-c_{2}\right)\right.\nonumber \\
 & -2bc_{3}\left(\frac{K^{2}}{u^{3}}-\frac{3}{u^{5}}\right)\cos v-\frac{3ac_{1}K}{u^{4}}\cos\left(v-s+c_{2}\right)\nonumber \\
 & -ac_{1}\left(\frac{K^{2}}{u^{3}}-\frac{3}{u^{5}}\right)\sin\left(v+s-c_{2}\right)-\frac{6bc_{3}K}{u^{4}}\sin v\nonumber \\
 & \left.+ac_{1}\left(\frac{K^{2}}{u^{3}}-\frac{3}{u^{5}}\right)\sin\left(v-s+c_{2}\right)\vphantom{\frac{\frac{1}{1}}{\frac{1}{1}}}\right]ds,\label{eq:ReLAMBDA3}\end{align}
 \begin{align}
\Im\Lambda_{1} & =\frac{a}{4}\int_{0}^{c}\left[\vphantom{\frac{\frac{1}{1}}{\frac{1}{1}}}\frac{3ac_{1}K}{u^{4}}\cos\left(v+2s-c_{2}\right)\right.\nonumber \\
 & -2bc_{3}\left(\frac{K^{2}}{u^{3}}-\frac{3}{u^{5}}\right)\cos\left(v+s\right)-\frac{6ac_{1}K\cos c_{2}}{u^{4}}\cos v\nonumber \\
 & +2bc_{3}\left(\frac{K^{2}}{u^{3}}-\frac{3}{u^{5}}\right)\cos\left(v-s\right)\nonumber \\
 & +\frac{3ac_{1}K}{u^{4}}\cos\left(v-2s+c_{2}\right)-\frac{6bc_{3}K}{u^{4}}\sin\left(v+s\right)\nonumber \\
 & -ac_{1}\left(\frac{K^{2}}{u^{3}}-\frac{3}{u^{5}}\right)\sin\left(v+2s-c_{2}\right)\nonumber \\
 & +2ac_{1}\cos c_{2}\left(\frac{K^{2}}{u^{3}}-\frac{3}{u^{5}}\right)\sin v+\frac{6bc_{3}K}{u^{4}}\sin\left(v-s\right)\nonumber \\
 & \left.-ac_{1}\left(\frac{K^{2}}{u^{3}}-\frac{3}{u^{5}}\right)\sin\left(v-2s+c_{2}\right)\vphantom{\frac{\frac{1}{1}}{\frac{1}{1}}}\right]ds,\label{eq:ImLAMBDA1}\end{align}
  \begin{align}
\Im\Lambda_{2} & =\frac{a}{4}\int_{0}^{c}\left[\vphantom{\frac{\frac{1}{1}}{\frac{1}{1}}}ac_{1}\left(\frac{K^{2}}{u^{3}}-\frac{3}{u^{5}}\right)\cos\left(v+2s-c_{2}\right)\right.\nonumber \\
 & +\frac{6bc_{3}K}{u^{4}}\cos\left(v+s\right)-\frac{6ac_{1}K\sin c_{2}}{u^{4}}\cos v\nonumber \\
 & -ac_{1}\left(\frac{K^{2}}{u^{3}}-\frac{3}{u^{5}}\right)\cos\left(v-2s+c_{2}\right)\nonumber \\
 & +\frac{6bc_{3}K}{u^{4}}\cos\left(v-s\right)+\frac{3ac_{1}K}{u^{4}}\sin\left(v+2s-c_{2}\right)\nonumber \\
 & -2bc_{3}\left(\frac{K^{2}}{u^{3}}-\frac{3}{u^{5}}\right)\sin\left(v+s\right)\nonumber \\
 & +2ac_{1}\sin c_{2}\left(\frac{K^{2}}{u^{3}}-\frac{3}{u^{5}}\right)\sin v\nonumber \\
 & -2bc_{3}\left(\frac{K^{2}}{u^{3}}-\frac{3}{u^{5}}\right)\sin\left(v-s\right)\nonumber \\
 & \left.-\frac{3ac_{1}K}{u^{4}}\sin\left(v-2s+c_{2}\right)\vphantom{\frac{\frac{1}{1}}{\frac{1}{1}}}\right]ds,\label{eq:ImLAMBDA2}\end{align}
  \begin{align}
\Im\Lambda_{3} & =\frac{b}{2}\int_{0}^{c}\left[\vphantom{\frac{\frac{1}{1}}{\frac{1}{1}}}ac_{1}\left(\frac{K^{2}}{u^{3}}-\frac{3}{u^{5}}\right)\cos\left(v+s-c_{2}\right)\right.\nonumber \\
 & +\frac{3bc_{3}K}{u^{4}}\cos v-ac_{1}\left(\frac{K^{2}}{u^{3}}-\frac{3}{u^{5}}\right)\cos\left(v-s+c_{2}\right)\nonumber \\
 & +\frac{3ac_{1}K}{u^{4}}\sin\left(v+s-c_{2}\right)-bc_{3}\left(\frac{K^{2}}{u^{3}}-\frac{3}{u^{5}}\right)\sin v\nonumber \\
 & \left.-\frac{3ac_{1}K}{u^{4}}\sin\left(v-s+c_{2}\right)\vphantom{\frac{\frac{1}{1}}{\frac{1}{1}}}\right]ds,\label{eq:ImLAMBDA3}\end{align}
 where $c_{1},$ $c_{2},$ and $c_{3}$ are defined in Eqs. (\ref{eq:c1-Def})
thru (\ref{eq:c3-Def}).  Insertion of Eq. (\ref{eq:par-RPSI-IPSI})
into Eqs. (\ref{eq:RE-gen}) and (\ref{eq:IE-gen}) yields the expression
given by  \[
\left(\begin{array}{c}
\Re\mathbf{E}\\
\Im\mathbf{E}\end{array}\right)=\pm\frac{\sigma E_{\textnormal{p}}}{a^{2}+b^{2}}\sum_{i=1}^{3}\left(\begin{array}{c}
\Re E_{i}\\
\Im E_{i}\end{array}\right)\mathbf{e}_{i},\]
 where \begin{align}
\left(\begin{array}{c}
\Re E_{i}\\
\Im E_{i}\end{array}\right) & =\gamma\omega\eta\left(\begin{array}{c}
\Im G_{i}\\
\Re G_{i}\end{array}\right)\nonumber \\
 & +\frac{g}{\omega}\sum_{l=1}^{3}\left[u_{l}u_{i}\left(\begin{array}{c}
\Im\Lambda_{l}\\
\Re\Lambda_{l}\end{array}\right)-\delta_{il}\left(\begin{array}{c}
\Im\Psi_{l}\\
\Re\Psi_{l}\end{array}\right)\right].\label{eq:RE-IE-vec}\end{align}
 Direct expansion of $\Re E_{i}$ and $\Im E_{i}$ for each $i=1,2,3$
to yields \begin{align}
\Re E_{1} & =\gamma\omega\eta\Im G_{1}\nonumber \\
 & +\frac{g}{\omega}\left(u_{1}^{2}\Im\Lambda_{1}+u_{2}u_{1}\Im\Lambda_{2}+u_{3}u_{1}\Im\Lambda_{3}-\Im\Psi_{1}\right),\label{eq:RE1}\end{align}
 \begin{align}
\Re E_{2} & =\gamma\omega\eta\Im G_{2}\nonumber \\
 & +\frac{g}{\omega}\left(u_{1}u_{2}\Im\Lambda_{1}+u_{2}^{2}\Im\Lambda_{2}+u_{3}u_{2}\Im\Lambda_{3}-\Im\Psi_{2}\right),\label{eq:RE2}\end{align}
 \begin{align}
\Re E_{3} & =\gamma\omega\eta\Im G_{3}\nonumber \\
 & +\frac{g}{\omega}\left(u_{1}u_{3}\Im\Lambda_{1}+u_{2}u_{3}\Im\Lambda_{2}+u_{3}^{2}\Im\Lambda_{3}-\Im\Psi_{3}\right),\label{eq:RE3}\end{align}
 and \begin{align}
\Im E_{1} & =\gamma\omega\eta\Re G_{1}\nonumber \\
 & +\frac{g}{\omega}\left(u_{1}^{2}\Re\Lambda_{1}+u_{2}u_{1}\Re\Lambda_{2}+u_{3}u_{1}\Re\Lambda_{3}-\Re\Psi_{1}\right),\label{eq:IE1}\end{align}
 \begin{align}
\Im E_{2} & =\gamma\omega\eta\Re G_{2}\nonumber \\
 & +\frac{g}{\omega}\left(u_{1}u_{2}\Re\Lambda_{1}+u_{2}^{2}\Re\Lambda_{2}+u_{3}u_{2}\Re\Lambda_{3}-\Re\Psi_{2}\right),\label{eq:IE2}\end{align}
 \begin{align}
\Im E_{3} & =\gamma\omega\eta\Re G_{3}\nonumber \\
 & +\frac{g}{\omega}\left(u_{1}u_{3}\Re\Lambda_{1}+u_{2}u_{3}\Re\Lambda_{2}+u_{3}^{2}\Re\Lambda_{3}-\Re\Psi_{3}\right).\label{eq:IE3}\end{align}
 The magnitude of $\mathbf{E}$ is given by \begin{align*}
E & =\sqrt{\mathbf{E}^{\dagger}\mathbf{E}}=\sqrt{\left(\Re\mathbf{E}-i\Im\mathbf{E}\right)\cdot\left(\Re\mathbf{E}+i\Im\mathbf{E}\right)}\\
 & =\sqrt{\Re\mathbf{E}\cdot\Re\mathbf{E}+\Im\mathbf{E}\cdot\Im\mathbf{E}}\end{align*}
  or with  Eq. (\ref{eq:RE-IE-vec}) substituted in for $\Re\mathbf{E}$
and $\Im\mathbf{E},$ I obtain \begin{equation}
\frac{E_{\textup{rad}}}{E_{\textnormal{p}}}\equiv\frac{E}{E_{\textnormal{p}}}=\frac{\sigma}{a^{2}+b^{2}}\left\{ \sum_{i=1}^{3}\left[\left(\Re E_{i}\right)^{2}+\left(\Im E_{i}\right)^{2}\right]\right\} ^{1/2},\label{eq:Erad-over-Ep}\end{equation}
 where $\Re E_{i}$ and $\Im E_{i}$ are defined in Eqs. (\ref{eq:RE1})
thru (\ref{eq:IE3}).

\section{Result}

\begin{figure}
\begin{centering}
\includegraphics[scale=0.35]{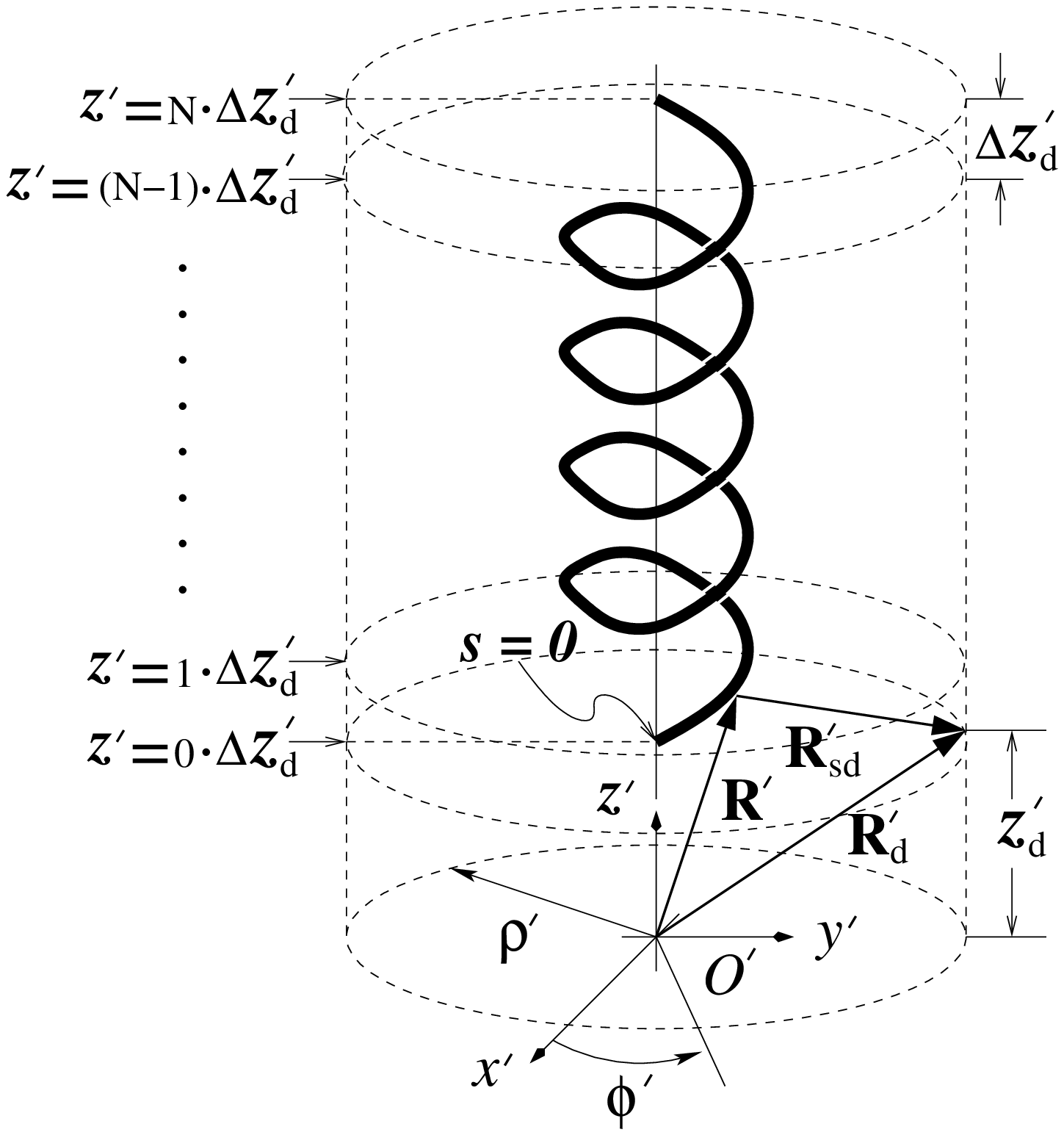}
\par\end{centering}

\caption{\label{CoF}Position of detector relative to the primed origin.}

\end{figure}

The fields are measured along the surface of cylindrical shell illustrated
in Fig.  \ref{CoF}. Relative to the $O^{\prime}$ frame of reference,
an arbitrary point on the surface of cylindrical shell is given by
\begin{equation}
\mathbf{R}_{\textup{d}}^{\prime}=\mathbf{R}^{\prime}+\mathbf{R}_{\textup{sd}}^{\prime},\label{eq:Rdp-Rp-plus-Rsdp}\end{equation}
 where $\mathbf{R}^{\prime}$ is from Eq. (\ref{eq:Rs}) and is given
by \begin{equation}
\mathbf{R}^{\prime}=a\cos\left(s\right)\mathbf{e}_{1}+a\sin\left(s\right)\mathbf{e}_{2}+bs\mathbf{e}_{3}\label{eq:R-prime-conf}\end{equation}
 with $\mathbf{e}_{1}\equiv\mathbf{x}^{\prime},$ $\mathbf{e}_{2}\equiv\mathbf{y}^{\prime},$
and $\mathbf{e}_{3}\equiv\mathbf{z}^{\prime}.$ In terms of cylindrical
coordinates $\left(\rho^{\prime},\phi^{\prime},z_{\textup{d}}^{\prime}\right),$
$\mathbf{R}_{\textup{d}}^{\prime}$ can be expressed as \[
\mathbf{R}_{\textup{d}}^{\prime}=\rho^{\prime}\cos\left(\phi^{\prime}\right)\mathbf{e}_{1}+\rho^{\prime}\sin\left(\phi^{\prime}\right)\mathbf{e}_{2}+z_{\textup{d}}^{\prime}\mathbf{e}_{3}\]
 and Eq. (\ref{eq:Rdp-Rp-plus-Rsdp}) may be solved for $\mathbf{R}_{\textup{sd}}^{\prime}$
to yield \begin{align}
\mathbf{R}_{\textup{sd}}^{\prime} & =\left(\rho^{\prime}\cos\phi^{\prime}-a\cos s\right)\mathbf{e}_{1}+\left(\rho^{\prime}\sin\phi^{\prime}-a\sin s\right)\mathbf{e}_{2}\nonumber \\
 & +\left(z_{\textup{d}}^{\prime}-bs\right)\mathbf{e}_{3},\label{eq:Rsd-prime}\end{align}
 where $\rho^{\prime}$ is a constant for a fixed cylindrical shell,
$\phi^{\prime}$ sweeps from $0$ to $2\pi,$ and $z^{\prime}=z_{\textup{d}}^{\prime}$
ranges from $-\infty$ to $\infty.$ 

\begin{figure}

\begin{centering}
\includegraphics[scale=0.35]{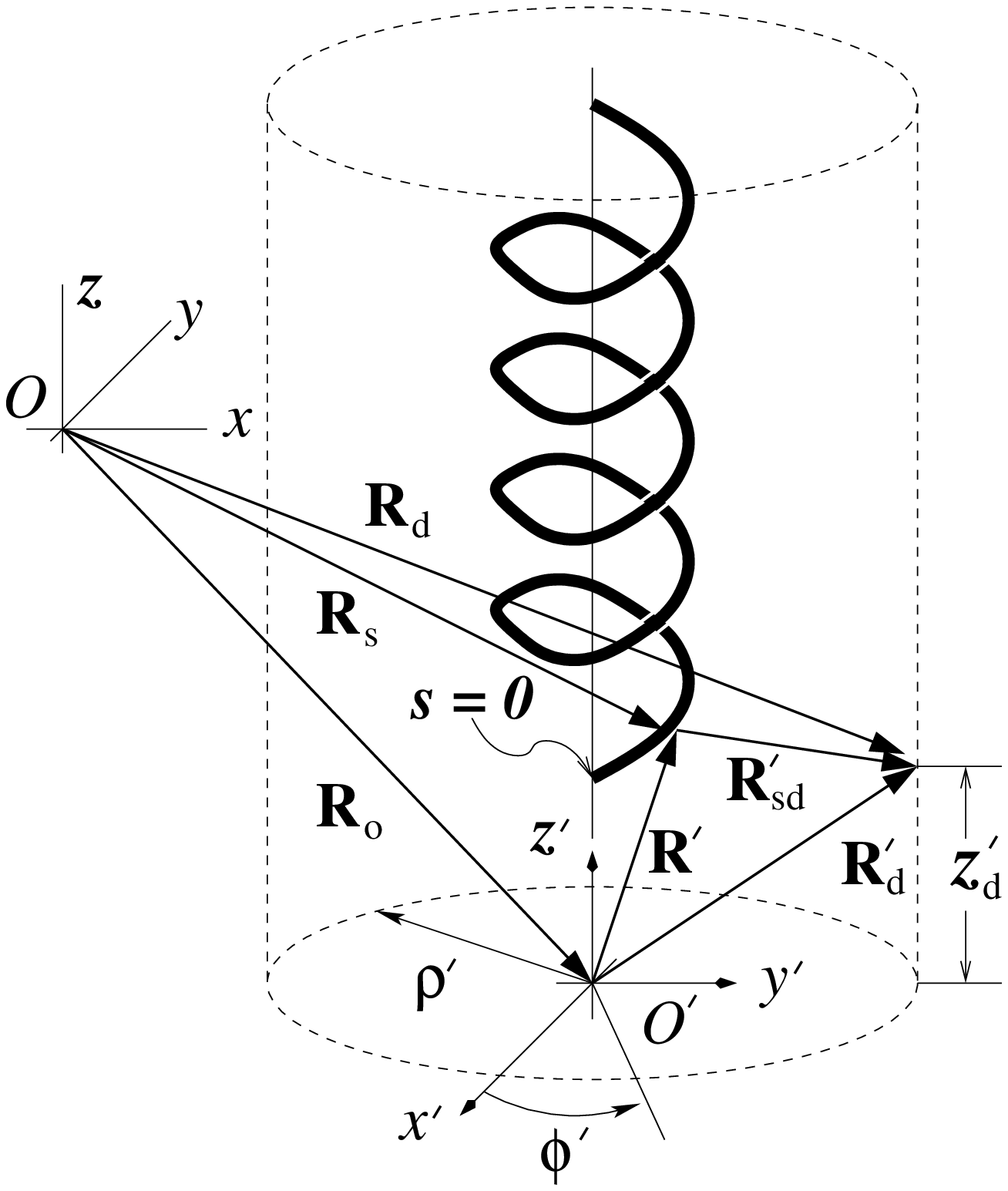}
\par\end{centering}

\caption{\label{DPR-O}Position of detector relative to the unprimed origin.}

\end{figure}

Relative to the frame of reference $O,$ Fig.  \ref{DPR-O} in which
frame the unit bases $x,y,z$ satisfy the condition $x\parallel x^{\prime},$
$y\parallel y^{\prime},$ $z\parallel z^{\prime},$  the locations
$\mathbf{R}_{\textup{d}}^{\prime}$ and $\mathbf{R}^{\prime}$ are
given by \[
\mathbf{R}_{\textup{d}}=\mathbf{R}_{\textup{o}}+\mathbf{R}_{\textup{d}}^{\prime},\quad\mathbf{R}_{\textup{s}}=\mathbf{R}_{\textup{o}}+\mathbf{R}^{\prime},\]
 where $\mathbf{R}_{\textup{s}}=\mathbf{R}_{\textup{o}}+\mathbf{R}^{\prime}$
is from Eq. (\ref{eq:Rs}). These relations can be combined with Eq.
(\ref{eq:Rdp-Rp-plus-Rsdp}) to yield \[
\mathbf{R}_{\textup{d}}=\mathbf{R}_{\textup{sd}}^{\prime}+\mathbf{R}^{\prime}+\mathbf{R}_{\textup{o}}=\mathbf{R}_{\textup{sd}}^{\prime}+\mathbf{R}^{\prime}+\sum_{i=1}^{3}x_{\textup{o}i}\mathbf{e}_{i}.\]
 With $\mathbf{R}^{\prime}$ inserted from Eq. (\ref{eq:R-prime-conf})
and $\mathbf{R}_{\textup{sd}}^{\prime}$ substituted in from Eq. (\ref{eq:Rsd-prime}),
$\mathbf{R}_{\textup{d}}$ becomes \begin{align}
\mathbf{R}_{\textup{d}} & =\left(x_{\textup{o}1}+\rho^{\prime}\cos\phi^{\prime}\right)\mathbf{e}_{1}+\left(x_{\textup{o}2}+\rho^{\prime}\sin\phi^{\prime}\right)\mathbf{e}_{2}\nonumber \\
 & +\left(x_{\textup{o}3}+z_{\textup{d}}^{\prime}\right)\mathbf{e}_{3}.\label{eq:Rd-pre-conf}\end{align}
 The $\mathbf{R}_{\textup{d}}$ in cylindrical coordinates can be
expressed as \begin{equation}
\mathbf{R}_{\textup{d}}=\rho\cos\phi\mathbf{e}_{1}+\rho\sin\phi\mathbf{e}_{2}+z_{\textup{d}}\mathbf{e}_{3},\label{eq:Rd}\end{equation}
 where $\rho$ is now the radius with respect to the unprimed reference
frame $O.$ Combining Eqs. (\ref{eq:Rd-pre-conf}) and (\ref{eq:Rd}),
I obtain \begin{align}
\left(\rho\cos\phi-x_{\textup{o}1}-\rho^{\prime}\cos\phi^{\prime}\right)\mathbf{e}_{1}\nonumber \\
+\left(\rho\sin\phi-x_{\textup{o}2}-\rho^{\prime}\sin\phi^{\prime}\right)\mathbf{e}_{2}\nonumber \\
+\left(z_{\textup{d}}-x_{\textup{o}3}-z_{\textup{d}}^{\prime}\right)\mathbf{e}_{3} & =0.\label{eq:condition-1}\end{align}
 Equation (\ref{eq:condition-1}) can only be satisfied if and only
if coefficients of $\mathbf{e}_{1},$ $\mathbf{e}_{2},$ and $\mathbf{e}_{3}$
are independently zero, hence \begin{alignat*}{1}
\rho\cos\phi-x_{\textup{o}1}-\rho^{\prime}\cos\phi^{\prime} & =0,\\
\rho\sin\phi-x_{\textup{o}2}-\rho^{\prime}\sin\phi^{\prime} & =0,\\
z_{\textup{d}}-x_{\textup{o}3}-z_{\textup{d}}^{\prime} & =0.\end{alignat*}
 The third relation readily gives \begin{equation}
z_{\textup{d}}=x_{\textup{o}3}+z_{\textup{d}}^{\prime}\label{eq:zd}\end{equation}
 and the first two relations rearranged to give \begin{alignat*}{1}
\rho\sin\phi & =x_{\textup{o}2}+\rho^{\prime}\sin\phi^{\prime},\\
\rho\cos\phi & =x_{\textup{o}1}+\rho^{\prime}\cos\phi^{\prime}.\end{alignat*}
 From the ratio of the two, i.e., $\tan\phi=\sin\phi/\cos\phi,$ I
obtain \begin{equation}
\phi=\arctan\left(\frac{x_{\textup{o}2}+\rho^{\prime}\sin\phi^{\prime}}{x_{\textup{o}1}+\rho^{\prime}\cos\phi^{\prime}}\right).\label{eq:phi}\end{equation}
 The $\rho$ is found by combining the two relations, i.e., $\rho\sin\phi+\rho\cos\phi,$
to get \[
\rho=\frac{x_{\textup{o}1}+x_{\textup{o}2}+\rho^{\prime}\left(\sin\phi^{\prime}+\cos\phi^{\prime}\right)}{\sin\phi+\cos\phi}.\]
 The $\sin\alpha+\cos\alpha$ can be combined utilizing Eq. (\ref{eq:sin-plus-cos})
to yield \begin{align*}
\sin\alpha+\cos\alpha & =\sqrt{2}\sin\left[\alpha+\arctan\left(1\right)\right]\\
 & =\sqrt{2}\sin\left(\alpha+\frac{\pi}{4}\right)\end{align*}
 and $\rho$ becomes \[
\rho=\frac{x_{\textup{o}1}+x_{\textup{o}2}+\rho^{\prime}\sqrt{2}\sin\left(\phi^{\prime}+\frac{\pi}{4}\right)}{\sqrt{2}\sin\left(\phi+\frac{\pi}{4}\right)}.\]
 Insertion of Eq. (\ref{eq:phi}) for $\phi$ yields the result%
\footnote{Notice that for the special case where $x_{\textup{o}1}=x_{\textup{o}2}=0,$
$\phi$ of Eq. (\ref{eq:phi}) reduces to $\phi=\arctan\left(\tan\phi^{\prime}\right)=\phi^{\prime}$
and the $\rho$ of Eq. (\ref{eq:rho}) becomes $\rho=\rho^{\prime}.$%
} \begin{equation}
\rho=\frac{x_{\textup{o}1}+x_{\textup{o}2}+\rho^{\prime}\sqrt{2}\sin\left(\phi^{\prime}+\frac{\pi}{4}\right)}{\sqrt{2}\sin\left[\arctan\left(\frac{x_{\textup{o}2}+\rho^{\prime}\sin\phi^{\prime}}{x_{\textup{o}1}+\rho^{\prime}\cos\phi^{\prime}}\right)+\frac{\pi}{4}\right]},\label{eq:rho}\end{equation}
 where $0\leq\phi^{\prime}<2\pi.$ With Eqs. (\ref{eq:zd}) and (\ref{eq:rho}),
the surface of cylindrical shell illustrated in Fig.  \ref{CoF} is
completely defined relative to the reference frame of $O.$  

In MKS system of units, where length is measured in meters, mass is
measured in kilograms, and time is measured in seconds, the constants
$g,$ $\eta,$ and $\gamma$ of Eqs. (\ref{eq:divB-FD}) thru (\ref{eq:curlB-FD})
are identified as \[
g=\frac{1}{4\pi\varepsilon_{\textup{o}}},\quad\eta=1,\quad\gamma=\frac{\mu_{\textup{o}}}{4\pi}\]
 and the constant $K$ in Eq. (\ref{eq:A-PDE}) gets identified as
\[
K=\omega\sqrt{\frac{\gamma\eta}{g}}=\omega\sqrt{\mu_{\textup{o}}\varepsilon_{\textup{o}}},\]
 where the free space electric permittivity $\varepsilon_{\textup{o}}$
and the magnetic permeability $\mu_{\textup{o}}$ have the value given
by \begin{align*}
\varepsilon_{\textup{o}} & \approx8.854\times10^{-12}\textnormal{ \textup{s}}^{4}\,\textnormal{\textup{A}}^{2}\,\textnormal{\textup{m}}^{-3}\,\textnormal{\textup{kg}}^{-1},\\
\mu_{\textup{o}} & =4\pi\times10^{-7}\textnormal{ \textup{m}}\,\textnormal{\textup{kg}}\,\textnormal{\textup{s}}^{-2}\,\textnormal{\textup{A}}^{-2}.\end{align*}

The total electric field and the total magnetic induction part of
the electromagnetic wave reaching the surface of an imaginary cylindrical
shell, illustrated in Fig. \ref{DPR-O}, can be summarized as \begin{eqnarray*}
\mathbf{E}_{\textup{total}}=\mathbf{E}_{\textup{P}}^{\prime}+\mathbf{E}_{\textup{rad}}, &  & \mathbf{B}_{\textup{total}}=\mathbf{B}_{\textup{P}}^{\prime}+\mathbf{B}_{\textup{rad}},\end{eqnarray*}
 where $\mathbf{E}_{\textup{P}}^{\prime}$ and $\mathbf{B}_{\textup{P}}^{\prime}$
respectively represent the electric field and the magnetic induction
parts of electromagnetic waves other than the radiation from nanohelix
reaching the detector. The measure of field amplification, therefore,
may be expressed as \begin{eqnarray*}
\frac{E_{\textup{total}}}{E_{\textup{P}}}=\frac{\left\Vert \mathbf{E}_{\textup{P}}^{\prime}+\mathbf{E}_{\textup{rad}}\right\Vert }{\left\Vert \mathbf{E}_{\textup{P}}\right\Vert }, &  & \frac{B_{\textup{total}}}{B_{\textup{P}}}=\frac{\left\Vert \mathbf{B}_{\textup{P}}^{\prime}+\mathbf{B}_{\textup{rad}}\right\Vert }{\left\Vert \mathbf{B}_{\textup{P}}\right\Vert }.\end{eqnarray*}
 In many situations, the magnitudes of $\mathbf{E}_{\textup{P}}^{\prime}$
and $\mathbf{B}_{\textup{P}}^{\prime}$ are only minute changes from
that of incidence wave, i.e., $\left\Vert \mathbf{E}_{\textup{P}}\right\Vert \approx\left\Vert \mathbf{E}_{\textup{P}}^{\prime}\right\Vert $
and $\left\Vert \mathbf{B}_{\textup{P}}\right\Vert \approx\left\Vert \mathbf{B}_{\textup{P}}^{\prime}\right\Vert ,$
and the previous expressions can be approximated as \begin{align}
\frac{E_{\textup{total}}}{E_{\textup{P}}} & \approx1+\frac{\left\Vert \mathbf{E}_{\textup{rad}}\right\Vert }{\left\Vert \mathbf{E}_{\textup{P}}\right\Vert }>\frac{\left\Vert \mathbf{E}_{\textup{P}}+\mathbf{E}_{\textup{rad}}\right\Vert }{\left\Vert \mathbf{E}_{\textup{P}}\right\Vert },\label{eq:Eto_over_Ep_done}\\
\frac{B_{\textup{total}}}{B_{\textup{P}}} & \approx1+\frac{\left\Vert \mathbf{B}_{\textup{rad}}\right\Vert }{\left\Vert \mathbf{B}_{\textup{P}}\right\Vert }>\frac{\left\Vert \mathbf{B}_{\textup{P}}+\mathbf{B}_{\textup{rad}}\right\Vert }{\left\Vert \mathbf{B}_{\textup{P}}\right\Vert }.\label{eq:Bto_over_Bp_done}\end{align}
 A successful rectification of ambient electromagnetic wave requires
the field amplification criterion defined in Eq. (\ref{eq:amplification_proof}),
\begin{eqnarray*}
\frac{E_{\textup{total}}}{E_{\textup{P}}}\gg1, &  & \frac{B_{\textup{total}}}{B_{\textup{P}}}\gg1,\end{eqnarray*}
 where the subscripts in the magnitudes of electric field and magnetic
induction have been modified from $\textup{rad}$ to $\textup{total}.$
Observing Eqs. (\ref{eq:Eto_over_Ep_done}) and (\ref{eq:Bto_over_Bp_done}),
the requirement for a successful rectification of ambient electromagnetic
wave is given by \begin{eqnarray*}
\frac{E_{\textup{rad}}}{E_{\textup{P}}}\gg1, &  & \frac{B_{\textup{rad}}}{B_{\textup{P}}}\gg1,\end{eqnarray*}
 which is the field amplification criterion defined in Eq. (\ref{eq:amplification_proof}). 

\begin{figure}[H]
\begin{centering}
\includegraphics[scale=0.7]{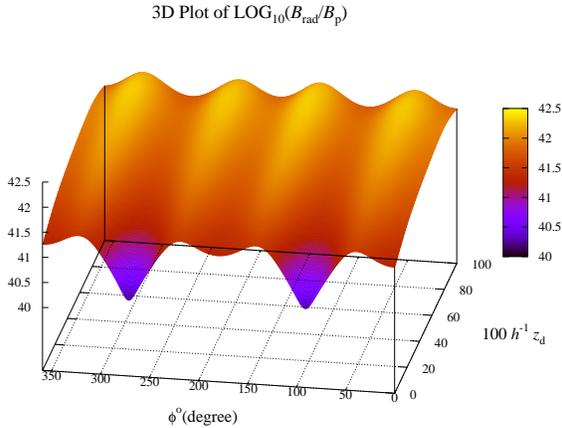}
\par\end{centering}

\caption{\label{B-field-3D} The ratio $B_{\textup{rad}}/B_{\textnormal{p}}$
measured on the surface of cylindrical screen of radius $\rho^{\prime}=273\,\textup{nm}.$ }

\end{figure}

\begin{figure}[H]
\begin{centering}
\includegraphics[scale=0.5]{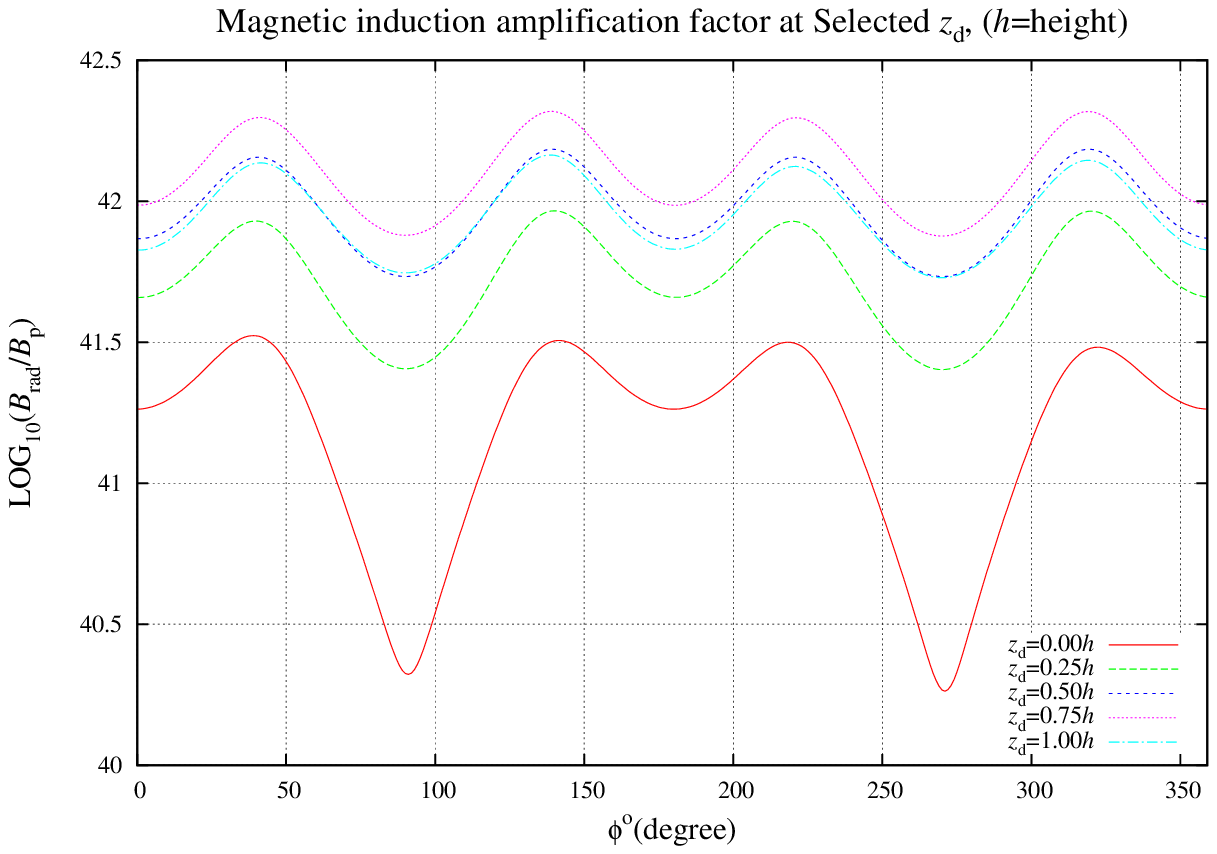}
\par\end{centering}

\caption{\label{B-field-1D} The ratio $B_{\textup{rad}}/B_{\textnormal{p}},$
which is measured on the surface of cylindrical screen of radius $\rho^{\prime}=273\,\textup{nm}$
and  sliced at indicated  $z_{\textup{d}}.$ }

\end{figure}

\begin{figure}[H]
\begin{centering}
\includegraphics[scale=0.7]{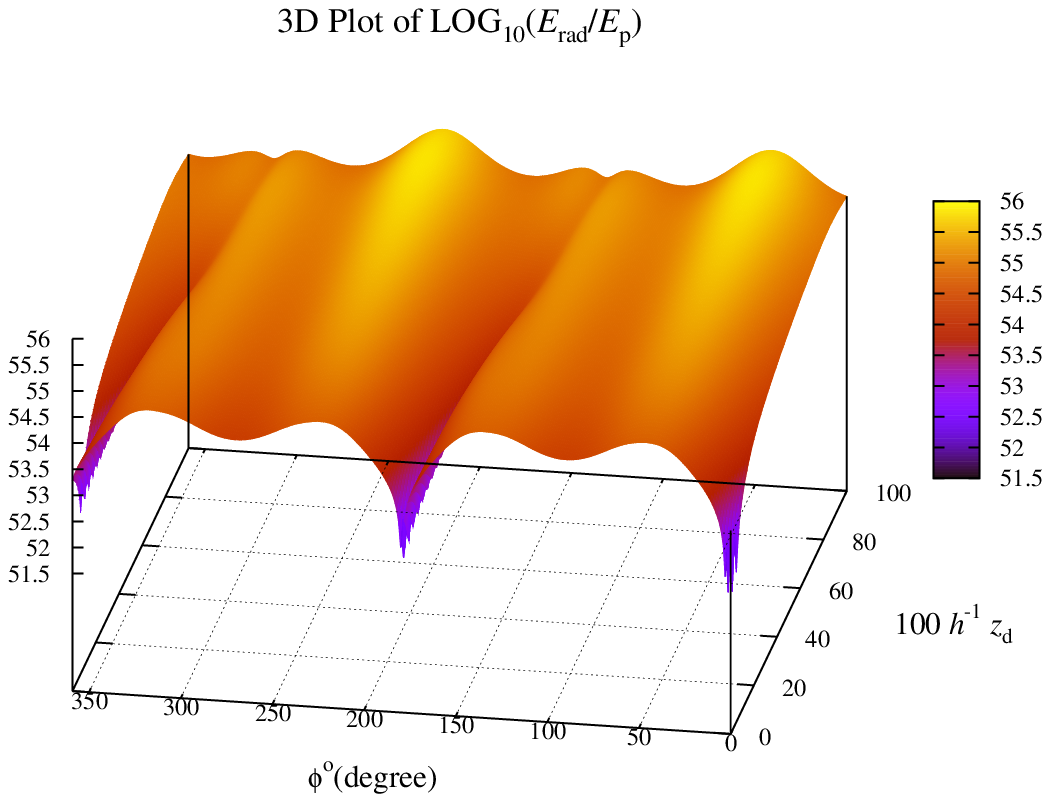}
\par\end{centering}

\caption{\label{E-field-3D} The ratio $E_{\textup{rad}}/E_{\textnormal{p}}$
measured on the surface of cylindrical screen of radius $\rho^{\prime}=273\,\textup{nm}.$ }

\end{figure}

\begin{figure}[H]
\begin{centering}
\includegraphics[scale=0.5]{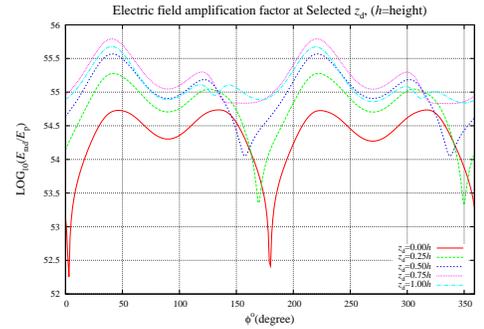}
\par\end{centering}

\caption{\label{E-field-1D} The ratio $E_{\textup{rad}}/E_{\textnormal{p}},$
which is measured on the surface of cylindrical screen of radius $\rho^{\prime}=273\,\textup{nm}$
and sliced at indicated  $z_{\textup{d}}.$  }

\end{figure}

I am now ready to plot the results. For convenience, the origins of
two reference frames, $O$ and $O^{\prime},$ were made to coincide
each other. This makes $x_{\textup{o}1}=x_{\textup{o}2}=0,$ $z_{\textup{d}}=z_{\textup{d}}^{\prime},$
and $\triangle z_{\textup{d}}=\triangle z_{\textup{d}}^{\prime}.$
Furthermore, it had been assumed that the helix winding started at
$z_{\textup{d}}=z_{\textup{d}}^{\prime}=0$ and the vacuum was assumed
for the medium holding both the finite helix and the propagating incidence
and radiated electromagnetic waves. That being said, Eqs. (\ref{eq:Brad-over-Bp})
and (\ref{eq:Erad-over-Ep}) are computed at the surface of cylindrical
screen of radius $\rho^{\prime},$ Eq. (\ref{eq:rho}), using Simpson
method coded in \textbf{FORTRAN 90} for numerical integration\citet{thomas-finney}
and assuming the following input values, \[
\textup{wavelength}\,\lambda=555.016\,\textup{nm},\]
 \[
\textup{wave number vector: }\mathbf{K}\left(K,0,0\right)=K\hat{\mathbf{x}}=\frac{2\pi}{\lambda}\hat{\mathbf{x}},\]
 \[
\textup{polarization: }\mathbf{E}_{\textup{P}}\left(\alpha_{1}=\frac{\pi}{2},\alpha_{2}=0,\alpha_{3}=\frac{\pi}{2}\right)=E_{\textup{P}}\hat{\mathbf{y}},\]
 \[
\textup{radius of helix: }a=40\,\textup{nm},\]
 \[
\textup{pitch of helix: }\textup{pitch}=50\,\textup{nm},\]
 \[
\textup{radius of screen: }\rho^{\prime}=273\,\textup{nm},\]
 \[
\textup{increment along z axis: }\triangle z_{\textup{d}}^{\prime}=0.0976\,\textup{nm},\]
 \[
\textup{helix conductivity: }\sigma=5\times10^{5}\,\textup{S},\]
 \[
\textup{fully streched length of helix: }l=5\,\textup{um},\]
 \[
\textup{helix winding number: }N_{\textup{w}}=19.5.\]
 The results are illustrated in Figs. \ref{B-field-3D} thru \ref{E-field-1D},
where  the parameter $h$ in Figs. \ref{B-field-3D} and \ref{E-field-3D}
represents the height of nanohelix, i.e., $h=bc,$ as illustrated
in Fig. \ref{HS-with-bc}. The results illustrated in Figs. \ref{B-field-1D}
and \ref{E-field-1D}, therefore, respectively represent slices of
Figs. \ref{B-field-3D} and \ref{E-field-3D} at helix height indicated
by $z_{\textup{d}}.$ 

\begin{figure}[H]
\begin{centering}
\includegraphics[scale=0.7]{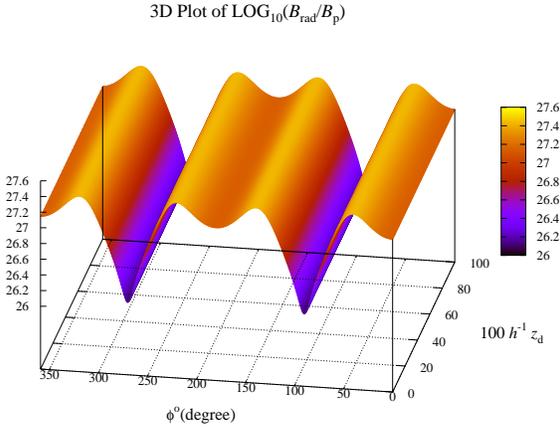}
\par\end{centering}

\caption{\label{B-field-3D_fat} The ratio $B_{\textup{rad}}/B_{\textnormal{p}}$
measured on the surface of cylindrical screen of radius $\rho^{\prime}=233.04\,\textup{um}.$ }

\end{figure}

\begin{figure}[H]
\begin{centering}
\includegraphics[scale=0.7]{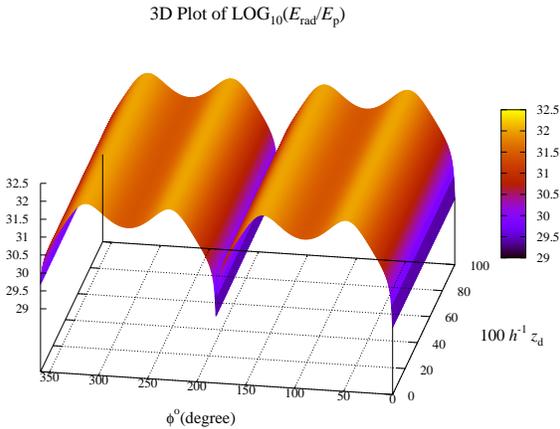}
\par\end{centering}

\caption{\label{E-field-3D_far} The ratio $E_{\textup{rad}}/E_{\textnormal{p}}$
measured on the surface of cylindrical screen of radius $\rho^{\prime}=233.04\,\textup{um}.$ }

\end{figure}

For the particular case where the input electromagnetic wave is specified
by the wave number vector $\mathbf{K}=K\hat{\mathbf{x}}$ and $\mathbf{E}_{\textup{P}}=E_{\textup{P}}\hat{\mathbf{y}},$
the resulting induced radiation from a single nanohelix is characterized
by an electric field part which is distorted in profile. The distortion
in electric field profile can be attributed to the local geometrical
configuration of the nanohelix, e.g., the winding pitch, etc. At distances
far away from nanohelix, such effects arising from the geometrical
configuration of nanohelix should be averaged out, resulting in a
radiated electric field profile which is symmetrical in shape. To
confirm such behavior for the scattered fields at distances far away
from scattering nanohelix, Eqs. (\ref{eq:Brad-over-Bp}) and (\ref{eq:Erad-over-Ep})
were re-computed with only the radius of cylindrical screen changed
to $\rho^{\prime}=233.04\,\textup{um},$ which is about thousand times
larger than the previous value for the radius, $\rho^{\prime}=273\,\textup{nm}.$
The results are shown in Figs. \ref{B-field-3D_fat} and \ref{E-field-3D_far}
for the ratios involving electric and magnetic fields, respectively.
As expected, the profiles for both magnetic induction and electric
field portray a symmetry about common axis. 

The proposed energy harvesting device based on nanohelices is actually
an array of many pixels, where the term pixel has been adopted from
display technology to denote a smaller energy harvesting device unit
composing of single nanohelix (or few nanohelices). In such many-body
(many-nanohelix) system, contributions arising from interaction with
different nanohelices must also be taken into account for a total
effect, which is a very challenging task. However, without having
to go through such difficult calculations involving many-body effect,
one can be sure of the existence of locations where constructive and
destructive interference of wavelets occur within the layer containing
nanohelices. The {}``number'' wins in energy harvesting device based
on nanohelices. Since there are billions of nanohelices in the system,
even if only one hundredth of them actually participate in the energy
harvesting, the total output power generated would be substantial.
And, the demonstrated calculation based on single nanohelix, supports
the possibility of harvesting energy utilizing nanohelices. Any calculations
involving many-nanohelices effect should be deferred for the optimization
stage of the development process.

\section{Concluding remarks}

The presented energy harvesting device based on nanohelices, in all
respects, can be thought of as a miniaturized version of rectifier
circuits with transformers found in many electronic systems. The only
difference is that rectifier based on nanohelices rectify ambient
electromagnetic waves, whereas the conventional rectifiers rectify
\textbf{AC} source from the household wall outlet. As with all rectifiers,
the rectification condition defined in Eq. (\ref{eq:rectification_condition})
must be satisfied before the proposed device can actually convert
ambient electromagnetic waves into a useful \textbf{DC} electrical
power. The rectification condition can be satisfied if the condition
defined in Eq. (\ref{eq:amplification_proof}) can be met, \begin{eqnarray*}
\frac{E_{\textup{rad}}}{E_{\textup{P}}}\gg1, &  & \frac{B_{\textup{rad}}}{B_{\textup{P}}}\gg1.\end{eqnarray*}
 In this work,  by utilizing the secondary radiation process, I have
explicitly shown that the condition imposed by Eq. (\ref{eq:amplification_proof})
becomes feasible with nanohelices.

\section{Acknowledgments}

The author acknowledges the support for this work provided by Samsung
Electronics, Ltd.

\end{document}